\documentclass[twocolumn]{pasj00}
\usepackage{enumerate}
\def\mbf#1{\mbox{\boldmath ${#1}$}}
\def\Alfven{Alfv\'{e}n~}

\SetRunningHead{Suzuki et al.}{Stellar Winds from Young Suns}
\Received{}
\Accepted{}
\Published{}

\begin{document}

\title{Saturation of Stellar Winds from Young Suns}

\author{Takeru K. Suzuki$^{1}$, Shinsuke Imada$^{2,4}$, Ryuho Kataoka$^{3}$, 
Yoshiaki Kato$^{4}$, Takuma Matsumoto$^{1}$, Hiroko Miyahara$^{5}$, 
Saku Tsuneta$^{4}$}
\email{stakeru@nagoya-u.jp}
\altaffiltext{1}{Department of Physics, Nagoya University,
Furo-cho, Chikusa, Nagoya, Aichi, 464-8602, Japan}
\altaffiltext{2}{Solar-Terrestrial Environmental Laboratory, Nagoya University,
Furo-cho, Chikusa, Nagoya, Aichi, 464-8601, Japan}
\altaffiltext{3}{Interactive Research Center of Science, Tokyo Institute of
Technology, 2-12-1 Ookayama, Meguro-ku, Tokyo, 152-8550 Japan}
\altaffiltext{4}{Solar and Plasma Astrophysics Division, National Astronomical 
Observatory, 2-21-1, Osawa, Mitaka, Tokyo, 181-8588, Japan}
\altaffiltext{5}{Institute for Cosmic Ray Research, The University of Tokyo,
5-1-5 Kashiwanoha, Kashiwa, Chiba 277-8582, Japan}



\maketitle

\begin{abstract}
We investigate mass losses via stellar winds from sun-like main sequence 
stars with a wide range of activity levels.
We perform forward-type magnetohydrodynamical 
numerical experiments for \Alfven wave-driven stellar winds with 
a wide range of the input Poynting flux from the photosphere.
Increasing the magnetic field 
strength and the turbulent velocity at the stellar photosphere from the 
current solar level, the mass loss rate rapidly increases at first 
owing to the suppression of the reflection of the \Alfven waves. 
The surface materials are lifted up by the magnetic pressure associated with 
the \Alfven waves, and the cool dense chromosphere is intermittently extended 
to 10 -- 20 \% of the stellar radius. 
The dense atmospheres enhance the radiative losses and eventually most of 
the input Poynting energy from the stellar surface escapes
by the radiation. As a result, there is no more sufficient energy 
remained for the kinetic energy of the wind; the stellar wind saturates 
in very active stars, as observed in Wood et al.  
The saturation level is positively correlated with $B_{r,0}f_0$, 
where $B_{r,0}$ and $f_0$ are the magnetic field strength and the filling 
factor of open flux tubes at the photosphere. 
If $B_{r,0}f_0$ is relatively large $\gtrsim 5$ G, the 
mass loss rate could be as high as 1000 times. 
If such a strong mass loss lasts for $\sim 1$ billion years, 
the stellar mass itself is affected, which could be a solution to 
the faint young sun paradox.
We derive a Reimers-type scaling relation that estimates the mass loss rate 
from the energetics consideration of our simulations. 
Finally, we derive the evolution of the mass loss 
rates, $\dot{M}\propto t^{-1.23}$, of our simulations, combining with 
an observed time evolution of X-ray flux from sun-like stars, 
which is shallower than 
$\dot{M}\propto t^{-2.33\pm 0.55}$
in Wood et al.(2005). 
\end{abstract}

\section{Introduction}
Main sequence stars with mass roughly below 1.5 solar mass ($M_{\odot}$ 
hereafter) posses a surface convective layer, which is the main origin 
of X-ray and wind activities from such solar-type stars; 
magnetic fields are generated in a surface convective layer by dynamo 
mechanisms 
(e.g., Choudhuri et al.1995; Brun et al.2004; Hotta et al.2012)
and turbulent motions 
associated with the magnetoconvection are the source to  
drive flares and outflows. Such magnetic activities are themselves 
interesting phenomena with fruitful physics, and in addition 
they may also affect planetary circumstances around central stars. 
Various researches have been carried out recently from this viewpoint 
(e.g., Terada et al.2009; Sterenborg et al.2011; Lammer et al.2012). 

Young sun-like stars are very active:
The observed X-ray flux is up to $\sim 1000$ times larger than 
the present solar level \citep{gud97,gud04}, and the X-ray temperature is 
also higher \citep{rib05,tel05}. 
There is an observational implication that a young star could 
have a very thick chromosphere extending to 10-20\% of the 
stellar radius \citep{cze12}, which is in contrast to the thin chromosphere 
of the present Sun with width of 0.1-1\% of the solar radius ($R_{\odot}$ 
hereafter).
These observations show that in young sun-like stars the atmospheric 
materials are lifted up to higher altitudes and heated up more intensely 
in upper regions.   
This is probably related to the strong magnetic fields with an order of 
kG or even larger observed in young main sequence stars  
(Donati \& Collier Cameron 1997; Saar \& Brandenburg 1999; Saar 2001; 
see also Donati \& Landstreet 2009 for recent review), which are 
much stronger than the average strength of 1-10 G of the present-day Sun.

On the other hand, 
the mass loss rates derived from the comparison between 
observations of near-by stars with the spectral types of G,K,M and spherical 
symmetric steady-state hydrodynamical simulations show a different trend 
(Wood et al.2002; 2005); 
after an increase of the mass loss rate with the increasing X-ray 
flux, it saturates at $\sim$ 100 times of the current solar level and 
even drops in some very active stars. 
An explanation of the saturation is due to the change of the magnetic topology. 
It is expected that the surface area of active stars is mostly covered 
by closed magnetic structure as a result of the strong magnetic fields 
\citep{sch02}. Then, the atmospheric gas is confined in closed loops
rather than streaming out from open flux tubes \citep{wod05,vid09}. 

In this paper, we propose an alternative and additional mechanism for
the saturation. We focus on the dynamics in 
open magnetic flux tubes and show by numerical simulations that the 
saturation of the stellar winds could take place in open flux tube regions 
because of an enhancement of the radiative losses in active stars. 
The construction of the paper is as follows: After briefly introducing 
the setup of the MHD simulations in \S \ref{sec:smst}, we summarize 
in \S \ref{sec:engfm} basic formulations for the analysis of the wind 
energetics in the later sections. 
\S \ref{sec:restav} presents the main results of 
the simulated stellar winds. In \S \ref{sec:dis} we extensively discuss 
our results in light of observations of stellar winds (\S \ref{sec:cmpobs}) 
and chromospheres (\S \ref{sec:evchr}). In particular, we directly compare 
the simulated winds with the observations by Wood et al.(2002; 2005) 
in \S \ref{sec:cmpobs}. In \S \ref{sec:fysp}, we discuss the results in 
terms of the faint young Sun paradox.

\section{Simulation Setup}
\label{sec:smst}

\subsection{Setup}

\begin{figure}
\begin{center}
\FigureFile(70mm,){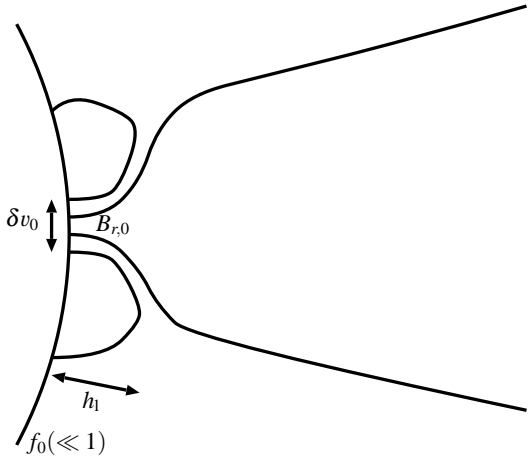}
\end{center}
\caption{Geometry of a flux tube and the input four parameters of the 
simulations.  $\delta v_0$ is the velocity amplitude of 
open field lines at the photosphere. Magnetic field strength, $B_{r,0}$, at 
the photosphere, a filling factor, $f_0$, of open flux tubes over the 
total photospheric surface, and a typical height, $h_{\rm l}$, of closed loops 
determine the properties of an open flux tube. $h_l$ corresponds to the 
location of the super-radial expansion of the flux tube. The combined variable, 
$(B_{r,0}f_0)$, determines the magnetic field strength in the outer region 
after the super-radial expansion finishes; $B_{r,0}f_0$ roughly corresponds to 
the large-scale field strength contributed from open flux tubes.  As described 
later in this paper, $\delta v_0$ and $B_{r,0}f_0$ determine the 
energy injection from the photosphere, and $h_{\rm l}$ affects the reflection 
of \Alfven waves in the chromosphere. }
\label{fig:kGp}
\end{figure}

We extend our one dimensional (1D) magnetohydrodynamical (MHD) simulation 
code originally developed for the present-day solar wind in
Suzuki \& Inutsuka (2005; 2006) to young active 
sun-like stars with strong magnetic fields and large velocity fluctuations 
at the photosphere (\S \ref{sec:inpp}).
In this paper, we focus on main sequence stars and fix the basic stellar 
parameters on those of the present-day Sun, mass, $M=M_{\odot}$,  
radius, $r_0=R_{\odot}$ and effective temperature, $T_{\rm eff} = 5780$ K. 
Compared to the current Sun, the radius of a 1$M_{\odot}$ star is 
slightly smaller by $\lesssim 10$ \% and the effective temperature also 
slightly lower within $\lesssim 200$ K at early epochs of the main sequence 
\citep{sac93}.
However, we use the same radius and effective temperature in our 
simulations for both active and inactive stars because we would like to study 
responses of the stellar winds to the changes of the magnetic fields and 
velocity fluctuations at the photosphere\footnote{Also, such subtle changes 
of the radius and effective temperature give almost negligible effects on 
the dynamics and the energetics of the stellar winds.}.

We dynamically solve ideal MHD equations with radiative cooling and thermal 
conduction in super-radially expanding flux tubes from the photosphere 
with density, $\rho_0=10^{-7}$g cm$^{-3}$ (e.g. \S 9 of Gray 
1992), and the sound speed $c_{\rm s,0}=6.31$ km s$^{-1}$ (derived from 
$T_{\rm eff}=5780$ K) at $r=r_0$ to the outer boundary at $r\approx 30 r_0$. 
We only consider the derivative with respect to $r$ but treat the 
three components of velocity and magnetic field.
We initially set up a static and cool ($T=10^4$ K) atmosphere and start 
simulations by injecting the three components of velocity fluctuations 
from the photosphere with amplitude of each component, $\delta v_0$, 
\begin{equation}
\langle \delta v_0^2 \rangle = \int_{\omega_{\rm min}}^{\omega_{\rm max}}P(\omega) 
d\omega, 
\end{equation}
where we adopt a frequency spectrum, $P(\omega)\propto \omega^{-1}$, from 
$1/\omega_{\rm max}=20$ seconds to $1/\omega_{\rm min}=30$ minutes. The 
normalization of $P(\omega)$ is determined to give $\delta v_0$.
Frequency spectrum of velocity perturbations have been 
observationally obtained by various instruments. Recent HINODE observation 
by \citet{mk10}, which 
obtained a frequency spectrum from 30 seconds to 70 minutes, shows a 
breaking power law with a flatter index=-0.6 in the lower frequency range, 
$1/\omega > 210$ seconds, and a steeper index = -2.4 in the higher frequency 
range. Our choice of $P(\omega) \propto \omega^{-1}$ is a reasonable fit to 
the observed spectrum with a single power law. For those who are interested in 
effects of different spectra, we would like to recall that \citet{szi06} 
performed the 1D MHD simulations with injected perturbations with white noise 
($P(\omega)\propto \omega^0$) and monochromatic waves of frequency $\omega_0$ 
($P(\omega)\propto \delta(\omega_0)$).

Various modes of MHD waves are excited by the surface fluctuations 
(e.g. Isobe et al.2008; Kato et al.2011), 
and propagate upwardly 
(e.g., Bogdan et al.2003; Okamoto \& De Pontieu 2011).
The \Alfven wave, among various modes, is supposed 
to play a major role in driving stellar winds mainly because it travels a long 
distance to the wind acceleration region 
(e.g., Alazraki \& Couturier 1971; Hollweg 1973).
A great advantage of the code is that we can determine 
mass loss rates as a direct output of the injected Poynting flux from the
photosphere.
Propagation, reflection, and dissipation of these MHD waves are 
directly treated by the dynamical MHD simulations. 
We setup the spatially variable grids in order to resolve the \Alfven waves 
with the highest frequency ($1/\omega_{\rm max}=20$ s.; 
shortest wavelength) we are injecting at least by $\gtrsim$ 10 grid points 
everywhere. 
On the other hand, we cannot handle scales which are smaller than 
the grid scale and assume that sub-grid-scale structures are instantly
transferred to the thermal energy by cascading processes or shocks to 
conserve the total energy equation. 
In our simulations, many shocklets form as a result of steepening 
of compressive waves which are nonlinearly generated from the propagating 
\Alfven waves (\S \ref{sec:engfm}). These shocklets with the scales smaller 
than the grid scale are assumed to dissipate and heat up surrounding gas in 
our treatment.

As for the cooling in the coronal region, we adopt the cooling table for the 
optically thin plasma with the solar abundance \citep{lm90,sd93}, in which 
the cooling rate, $q_{\rm R}$ (erg cm$^{-3}$s$^{-1}$), is 
proportional to $\rho^2$. For the cooling in the chromosphere, we adopt an 
empirical cooling rate $q_{\rm R}=4.5\times 10^9\rho$ erg cm$^{-3}$s$^{-1}$ by 
\citet{aa89}. This cooling rate takes into account an effect of optically 
thick cooling under the non-LTE conditions (but see Carlsson \& Leenaarts 
2012 for a more detailed treatment based on snapshots of 2D MHD simulations).
We smoothly connect the cooling rates in these two regimes by an interpolation.
In the transition region, the cooling rate is initially 
$\propto \rho$ at lower heights but shifted to $\rho^2$ at higher altitudes. 

We do not explicitly take into account stellar rotation in the dynamics of 
the stellar winds. 
If a star rotates by more than $\sim$ 20 times faster than the
present Sun, the terminal velocity of the wind will be affected (\S 
\ref{sec:strot}). 
However, the mass loss rate 
is affected little \citep{bm76}.
The stellar rotation rather significantly affects 
the strengths of the generated magnetic fields \citep{hj07}. 
In this paper, we consider this effect by incorporating wide ranges of 
the parameters on the magnetic fields. 

\subsection{Input Parameters}
\label{sec:inpp}
We investigate how the properties of the stellar winds depend on  
the following four parameters (Figure \ref{fig:kGp}): 
Velocity perturbation, $\delta v_0$, and 
radial magnetic field strength, $B_{r,0}$, a filling 
factor, $f_0$, of open\footnote{We call an `open' flux tube if the magnetic 
fields do not close in the simulation region.} flux tubes, 
which are all measured at the photosphere, and a typical height, $h_{\rm l}$, 
of closed loops, which surround an open flux tube we are considering. 
$h_{\rm l}$ determines the location of the rapid expansion of a flux tube 
(Figure \ref{fig:kGp}). 
From $f_0$ and $h_{\rm l}$, we adopt a functional form of a filling 
factor\footnote{In many literatures including ours 
(Suzuki \& Inutsuka 2005; 2006) a super-radial expansion factor, instead 
of a filling factor, is used to set up open flux tubes. 
Defining $f'(r)$ as a super-radial expansion factor, these two factor are 
simply related as $f(r) = f_0 f'(r)$. 
} which depends on 
$r$, 
\begin{equation}
f(r) = \frac{e^{\frac{r-r_0-h_{\rm l}}{h_{\rm l}}} + f_0 - (1-f_0)/e}
{e^{\frac{r-r_0-h_{\rm l}}{h_{\rm l}}}+1}.
\end{equation}
This is the same functional 
form as in \citet{kh76} with adopting $h_{\rm l}=R_1-R_{\odot}=\sigma$ 
in Equation (11) of their paper.
The conservation of magnetic flux fixes radial magnetic field, 
\begin{equation}
B_r = B_{r,0} \frac{f_0 r_0^2}{f(r)r^2}
\label{eq:magflx}
\end{equation}
We would like to note that the combined variable, $(B_{r,0}f_0)$,
determines the magnetic field strength in the outer region where the 
super-radial expansion already finishes, and corresponds to the field 
strength of larger-scale open flux tubes. 

\begin{table}
\begin{center}
{\footnotesize
\begin{tabular}{|c|c| |c|c|}
\hline
Label & $B_{r,0}f_0$(G) & Label & $B_{r,0}$(kG)\\
\hline
\hline
A & 0.3125 & a & 0.5\\
\hline
B & 0.625 & b & 1\\
\hline
C & 1.25 & c & 2\\
\hline
D & 2.5 & d & 4\\
\hline
E & 5 & e & 8\\
\hline
F & 10 & f & 16\\
\hline
\hline
Label & $\delta v_0$(km s$^{-1}$) & Label & $h_{\rm l}(/r_0)$\\
\hline
\hline
-2 & 0.669 & $\alpha$ & 0.01 \\
\hline
0 & 1.34 & $\beta$ & 0.03 \\
\hline
+1 & 1.89 & $\gamma$ & 0.1 \\
\hline
+2 & 2.68 & & \\
\hline
+3 & 3.79 & & \\
\hline
+4 & 5.35 & & \\
\hline
+5 & 7.57 & & \\
\hline
\end{tabular}
}
\end{center}
\caption{Labels for input parameters. For example, the label for the standard 
case is  `Cb0$\alpha$'.
\label{tab:prmlbl}
}
\end{table}

As a reference case for the present-day solar wind, we use the following 
values, which explain observed properties of polar coronal holes: 
$\delta v_0 = 1.34$ km s$^{-1}$, $B_{r,0} = 1$ kG, $f_0=1/800$, and 
$h_{\rm l}=0.01 R_{\odot}$($=7\times 10^3$ km).
The value of $\delta v_0$ is within the range of observed granulations 
\citep{hgr78,mk10}. 
We adopt the values of $B_{r,0}$ and $f_0$ 
to explain recent observations by HINODE \citep{tsu08,shi09,ito10,shi12}. 
In these papers, 
the authors reported that there are a number of super-radially open flux 
tubes anchored from strong magnetic field patches with $B_{r,0}\sim$ kG 
at the photosphere in polar regions. 
Comparing the photospheric magnetic field strength ($\sim 1$ kG) and measured 
interplanetary field strengths normalized at the earth orbit 
($\sim 1-10$ nT $=10^{-5}-10^{-4}$ G; e.g., Smith \& Balogh 2008), a 
typical filling factor, $f_0$, of open field regions can be estimated 
as an order of 1/1000 (see Equation \ref{eq:magflx}).

We perform 163 runs in wide ranges of the parameters: 
$\delta v_0 = (0.669 - 7.57)$ km s$^{-1}$, $B_{r,0}=(0.5 - 16)$ kG, 
$f_0 = (1/400 - 1/6400)$, and $h_{\rm l} = (0.01 - 0.1) r_0$,  
Compared with the parameters for the present-sun case just 
described above, we consider more cases with larger $\delta v_0$, 
larger $B_{r,0}$, smaller $f_0$, and larger $h_{\rm l}$.
We na\"{i}vely expect that $\delta v_0$ is 
larger in active stars, whereas too strong magnetic fields might inhibit 
footpoint motions of flux tubes \citep{kt05}. 
Since the actual $\delta v_0$ is quite uncertain observationally, we adopt 
the wide range $\delta v_0=0.669 - 7.57$ km s$^{-1}$, 
from $\sim10$\% of the sound speed (=6.31 km s$^{-1}$) to a moderately 
super-sonic regime at the photosphere.  
The photospheric magnetic field strengths, $B_{r,0}$, of active stars are 
expected to be larger than that of the present solar ({\it e.g.}, 
Donati et al. 2009). We should note that $B_{r,0}$ does not coincide with 
large-scale magnetic field strengths observed in solar-type stars 
({\it e.g.}, Figure 3 of Donati et al.2009), which will be discussed in more 
detail in \S \ref{sec:tmevl}. The large-scale field strength 
contributed from open flux tubes roughly corresponds to $B_{r,0}f_0$. 
In addition, the contribution from large-scale closed magnetic loops should 
be taken into account in order to estimate the total large-scale magnetic field 
strength. 
In this context, a filling factor, $f_0$, of open flux tubes 
is also an important parameter. At the same time, however, it is quite 
difficult to observationally determine typical $f_0$ of stars except for 
the sun. 
During the solar maximum of the present-day sun, a larger fraction of the 
surface is covered by closed structures \citep{hak05}, which implies $f_0$ 
is smaller. 
The magnetic states of active stars are considered to be a more extreme state 
of the solar maximum, and hence, we infer that $f_0$ of active stars is 
smaller. Thus, we consider the larger number of cases with smaller $f_0$. 
A typical height, $h_{\rm l}$, of closed loops of some active stars can be 
inferred from stellar flares through the RTV scaling law \citep{ros78}. 
\citet{sy99} derived a scaling relation for solar and stellar flares 
showing that the loop lengths of active stars are longer than those of the 
present sun. 

We label the simulated models by (A -- F) for $B_{r,0}f_0$, (a -- f) for 
$B_{r,0}$, ($-2$ -- $+5$) for $\delta v_0$, and ($\alpha$ -- $\gamma$) for 
$h_{\rm l}$, which are tabulated\footnote{The 
labels for $\delta v_0$ are the twice of the power-law indices of 2 
scaled by $\delta v_0=1.34$ km s$^{-1}$ of the standard case. 
For example, `+3' for $\delta v_0=3.79$ km s$^{-1}$ is from $(3.79_{\ldots})^2
= 2^{+3} \times (1.34_{\ldots})^2$ or $3.79_{\ldots}  = 2^{+3/2} \times 
1.34_{\ldots} $.} 
in Table \ref{tab:prmlbl}. Using these labels, for instance
the standard case ($B_{r,0}f_0=1.25$ G, $B_{r,0}=1$ kG, 
$\delta v_0=1.34$ km s$^{-1}$, and $h_{\rm l}=0.01r_0$) is labeled as 
`Cb0$\alpha$'.  

\section{Energetics Formulation}
\label{sec:engfm}

In this paper, we investigate how the mass losses are controlled by 
the surface properties by examining the energetics of the simulated 
stellar winds. 
We firstly consider the time-averaged structure of each simulation run 
in the next section (\S \ref{sec:restav}). In this section we summarize 
some basic equations for the wind energetics. 
Under the steady state conditions, an equation of the total energy can 
be written as (e.g., Fisk et al.1999; Suzuki 2006) 
$$
\hspace{-3cm}
\mbf{\nabla \cdot}\left[\rho \mbf{v} \left(\frac{v^2}{2} + \frac{\gamma}
{\gamma -1}\frac{p}{\rho} - \frac{GM}{r}\right)\right.
$$
\begin{equation}
\label{eq:engeq1}
\left. - \frac{1}{4\pi}\mbf{(v\times
B)\times B} +\mbf{F_{\rm c}}\right] + q_{\rm R} = 0 ,
\end{equation}
where $\mbf{F_{\rm c}}$ 
is thermal conductive flux, 
and the other variables have the conventional meanings. 
We focus on the radial component of the equation from now. We can write 
the divergence of an arbitrary vector, $\mbf{A}$, as
\begin{equation}
\mbf{\nabla \cdot A} = \frac{1}{r^2 f}\frac{\partial}{\partial r}(r^2 f A_r)
\end{equation}
to take into account the super-radial expansion of an open flux tube. 
The integration of Equation (\ref{eq:engeq1}) from 
an arbitrary reference point, $r_{\rm ref}$, to $r$ gives  
$$
\hspace{-2cm}4\pi r^2 f\left[\rho v_r\left(\frac{v^2}{2} 
+ \frac{\gamma}{\gamma -1}\frac{p}{\rho} - \frac{GM}{r}\right)
+ v_r\frac{B_{\perp}^2}{4\pi}\right.
$$
$$
\left. -B_r\frac{v_{\perp}B_{\perp}}{4\pi} + F_{{\rm c},r} \right] 
+ 4\pi\int_{r_{\rm ref}}^r q_{\rm R}r^2f dr
$$
$$
\hspace{-2cm}=\dot{M}\left(\frac{v^2}{2} + \frac{\gamma}{\gamma -1}
\frac{p}{\rho} + \frac{B_{\perp}^2}{4\pi\rho} - \frac{GM}{r}\right) 
$$
\begin{equation}
\label{eq:engeq2}
- \Phi_B\frac{v_{\perp}B_{\perp}}{4\pi} + 4\pi r^2f F_{{\rm c},r} 
+ 4\pi\int_{r_{\rm ref}}^r q_{\rm R}r^2f dr = {\rm const.}, 
\end{equation}
where subscript, $\perp$, indicates the perpendicular components to 
the radial direction, 
$\dot{M}$ is mass loss rate, 
\begin{equation}
\dot{M} = 4\pi r^2 f \rho v_r, 
\end{equation}
and $\Phi_B$ is radial magnetic flux, 
\begin{equation}
\Phi_B =  4\pi r^2 f B_r. 
\end{equation}

The energy flux of \Alfven waves along with $r$ direction can be written as 
\begin{equation}
F_{\rm A} = v_r \left(\rho\frac{v_{\perp}^2}{2} + \frac{B_{\perp}^2}{4\pi}\right)
- B_r\frac{v_{\perp}B_{\perp}}{4\pi},
\label{eq:Alfflx}
\end{equation}
\citep{jac77,cra07}.
One may notice that $F_{\rm A}$ partly consists of Equation (\ref{eq:engeq2}). 
Integrating over the total area of the open flux tubes on a stellar surface, 
we can define \Alfven wave luminosity, 
$L_{\rm A}$, as 
\begin{eqnarray}
L_{\rm A}(r)f(r) &=& 4\pi r^2 f(r) F_{\rm A}(r) \\ \nonumber
&=& \dot{M}\left(\frac{v_{\perp}^2}{2} + \frac{B_{\perp}^2}{4\pi\rho} \right) 
- \Phi_B\frac{v_{\perp}B_{\perp}}{4\pi} .
\label{eq:Alflum}
\end{eqnarray}
Since we consider the energy transfer in super-radially open flux tubes, 
luminosity, $L_{\rm A}f$, from open flux tubes, instead of $L_{\rm A}$ 
integrated over the total area, should be treated in the energy conservation.

The right-hand side of Equation (\ref{eq:Alfflx}) is separated into the 
two parts; 
the first term indicates 
the energy flux advected by background flow, $v_r$, and the second term, which 
exists even in static media, is the Poynting flux involving magnetic tension.  
We can introduce Els\"{a}sser variables, 
\begin{equation}
z_{\pm} = v_{\perp} \mp \frac{B_{\perp}}{\sqrt{4\pi\rho}},
\label{eq:elsvar}
\end{equation}
where $z_{+}(z_{-})$ denotes 
an amplitude of \Alfven waves which travels to the (anti-)parallel 
direction with $B_r$.
Using $z_{\pm}$, the second term of Equation (\ref{eq:Alfflx}) can be 
expressed as
\begin{equation}
-B_r\frac{v_{\perp}B_{\perp}}{4\pi} = \frac{1}{4}\rho(z_{+}^2 - z_{-}^2) v_{\rm A}, 
\end{equation}
where $v_{\rm A}=\frac{B_r}{\sqrt{4\pi\rho}}$ is \Alfven velocity. 
This expression illustrates that $\frac{1}{4}\rho z_{+}^2$ 
($\frac{1}{4}\rho z_{-}^2$ ) corresponds to the energy density of 
\Alfven waves to the (anti-)parallel direction with $B_r$, and 
that $-B_r\frac{v_{\perp}B_{\perp}}{4\pi}$ corresponds to the net positively 
traveling (parallel with $B_r$) Poynting flux associated with \Alfven waves 
in static media.

After the injection of \Alfven waves from the photosphere $L_{\rm A}f$ 
decreases because of various types of damping processes, such as turbulent 
cascade \citep{mat99,cha05,cra07,vv07}, phase mixing \citep{hp83,gra00,dem00}, 
nonlinear mode conversion to compressive waves by magnetic pressure, 
$B_{\perp}^2/8\pi$
(ponderomotive force; Kudoh \& Shibata 1999; Suzuki \& Inutsuka 2005; 2006) 
and by parametric decay \citep{gol78,ter86,szi06,nar09}, 
ioncyclotron resonance (e.g., Axford \& McKenzie 1997; Tu \& Marsch 2001).
Because of the 1D MHD approximation our simulations cannot take 
into account all these processes but mainly consider the nonlinear 
mode conversion to compressive waves by the ponderomotive force and parametric 
decay. 
However, our simulations are calibrated to reproduce the current solar wind, 
namely this can give a reasonable heating profile at least for the solar 
wind observed today although 3D and/or kinetic effects which we do not cover 
might play a role in the actual wave dissipation. We extend 
the calibrated case to extreme conditions for young suns. Recently,
\citet{ms12} have performed 2D MHD simulations, which can handle phase 
mixing and a part of turbulent cascade in addition to the nonlinear generation 
of compressive waves. Interestingly, the 1D \citep{szi05,szi06} and the 2D 
\citep{ms12} simulations give the similar wind structures for the 
present-day solar wind although the dissipation channels are different.  
In future works, we extend the 2D MHD simulation to the conditions for 
young active suns.

After the damping of the \Alfven waves, a small fraction of the input energy 
is finally transferred to the kinetic energy of stellar winds,  
\begin{equation}
L_{\rm K}(r)f(r) = \dot{M}\frac{v_r^2}{2} = 4\pi r^2 f(r)\rho v_r \frac{v_r^2}{2}
=4\pi r^2 f(r) F_{\rm K}(r),
\end{equation}
where $F_{\rm K}$ and $L_{\rm K}$ are kinetic energy flux and luminosity. 

Our main aim is to understand how the wind kinetic energy luminosity, 
$L_{\rm K}(r_{\rm out})f(r_{\rm out})=L_{\rm K}(r_{\rm out})$ 
(note $f(r_{\rm out})=1$), at the outer boundary, $r_{\rm out}\approx 30 r_0$, 
is determined by the 
input Poynting flux $(L_{\rm A}f)_0$ at the photosphere. Hereafter, 
we express, {\it e.g.} $(L_{\rm A}f)_0$ and $L_{\rm K,out}$, {\it etc.} for 
$L_{\rm A}(r_0)f_0$ and $L_{\rm K}(r_{\rm out})$ for simplicity. 
In our simulations, we only inject velocity perturbations with 
$\delta v_0$ from the photosphere without magnetic field perturbation. 
Then, the input energy flux can be written as $F_{\rm A,0} = \rho_0 
\langle \delta v_0^2\rangle v_{\rm A,0}$ in the static background ($v_r=0$), 
where $\langle \delta v_0^2\rangle$ is kinetic energy per mass of the sum of 
the two transverse components.
Then, the input wave luminosity is 
\begin{equation}
(L_{\rm A}f)_0 = 4\pi r_0^2 f_0\rho_0 \langle \delta v_0^2\rangle 
v_{\rm A,0}.
\label{eq:LA0}
\end{equation}

\begin{figure}
\begin{center}
\FigureFile(81mm,){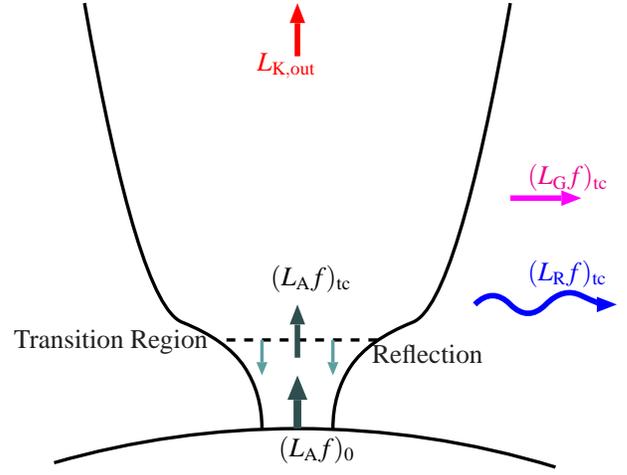}
\end{center}
\caption{Schematic picture for the energetics in an open flux tube. Please see 
{\it text} for the notation of each variable.}
\label{fig:sweng}
\end{figure}

We examine the energetics of the stellar winds by two steps (Figure 
\ref{fig:sweng}); we firstly 
inspect the energetics in the chromospheres. Here, the reflection of outgoing
\Alfven waves play an important role because the \Alfven speeds change rapidly 
because of the steep decrease of the densities.  
We define $r=r_{\rm tc}$ as the top of the chromosphere where 
$T=2\times 10^4$ K. The density, $\rho_{\rm tc}$, at $r=r_{\rm tc}$ 
is determined by the energy balance among the heating by wave dissipation, 
the downward thermal conduction, and the radiative cooling in the transition 
region \citep{ros78}. For larger heating, $\rho_{\rm tc}$ is larger because of 
the larger chromospheric evaporation through the downward thermal conduction 
as a result of the larger heating in the corona.
We measure $(L_{\rm A}f)_{\rm tc}$ at $r=r_{\rm tc}$ to quantify the 
reflection (\S \ref{sec:rlrtr}), and inspect the transmissivity,  
\begin{equation}
c_{\rm T} = (L_{\rm A}f)_{\rm tc} / (L_{\rm A}f)_0 , 
\label{eq:trnsm}
\end{equation}
of the \Alfven waves through the chromosphere in different cases. 
 

Next, we examine how much fraction of the surviving wave energy, 
$(L_{\rm A}f)_{\rm tc}$, 
is finally transferred to the wind kinetic energy (\S \ref{sec:rgrtr}).
By comparing dominant terms at $r_{\rm tc}$ and at 
$r_{\rm out}$ in Equation (\ref{eq:engeq2}), we can derive an energy  
conservation relation:
\begin{equation}
L_{\rm K,out} \approx (L_{\rm A}f)_{\rm tc} - (L_{\rm R}f)_{\rm tc}
- (L_{\rm G}f)_{\rm tc}, 
\label{eq:engsw1}
\end{equation}
where 
\begin{equation}
(L_{\rm R}f)_{\rm tc} \equiv 4\pi\int_{r_{\rm tc}}^{r_{\rm out}} q_{\rm R}r^2f dr
\end{equation}
is the radiation loss from $r=r_{\rm tc}$ to the outer region, 
and the gravitational loss,   
\begin{equation}
(L_{\rm G}f)_{\rm tc} \equiv \dot{M}\frac{GM}{r_{\rm tc}}. 
\end{equation}
Please note that the energy loss by the downward thermal conduction in the
transition regions is included in the radiation loss term, 
$(L_{\rm R}f)_{\rm tc}$, because the thermal conductive flux from the upper 
coronae mainly escapes by the radiation in the transition regions 
\citep{ros78}. As explained previously in this section, the 
density, $\rho_{\rm tc}$, at $r_{\rm tc}$, which corresponds to the `base' 
density for stellar winds, is determined by this energy balance.

\section{Results}
\label{sec:restav}

\subsection{Photosphere -- Wind connection}
\label{sec:pwcon}

\begin{figure}
\begin{center}
\FigureFile(90mm,){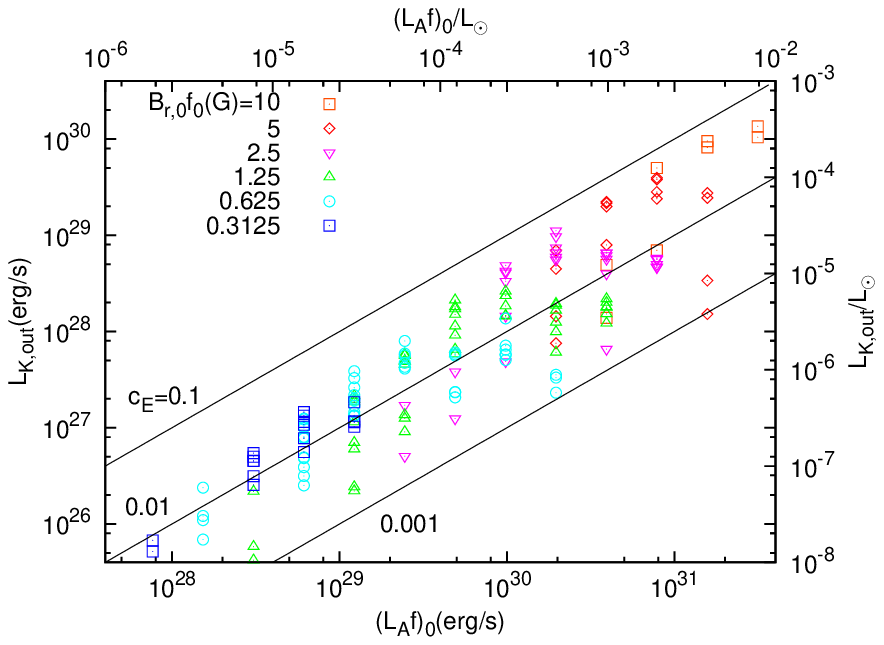}
\FigureFile(85mm,){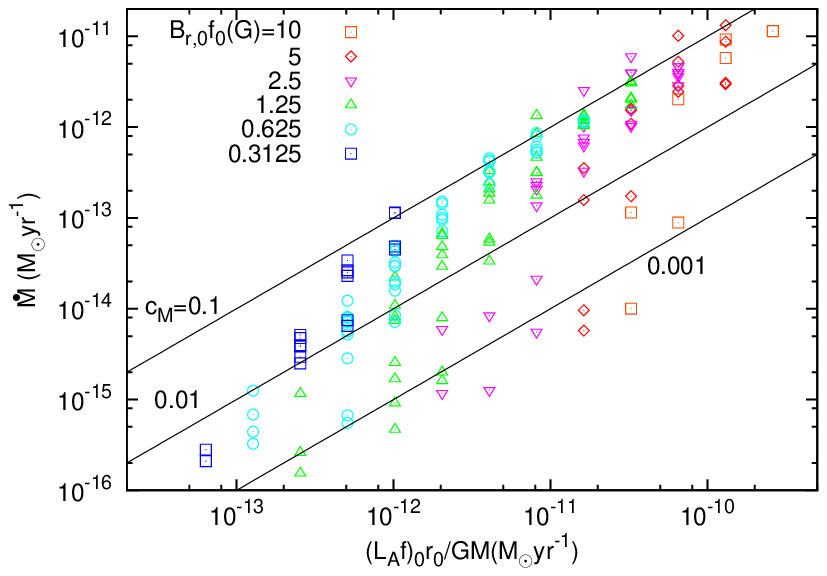}
\FigureFile(85mm,){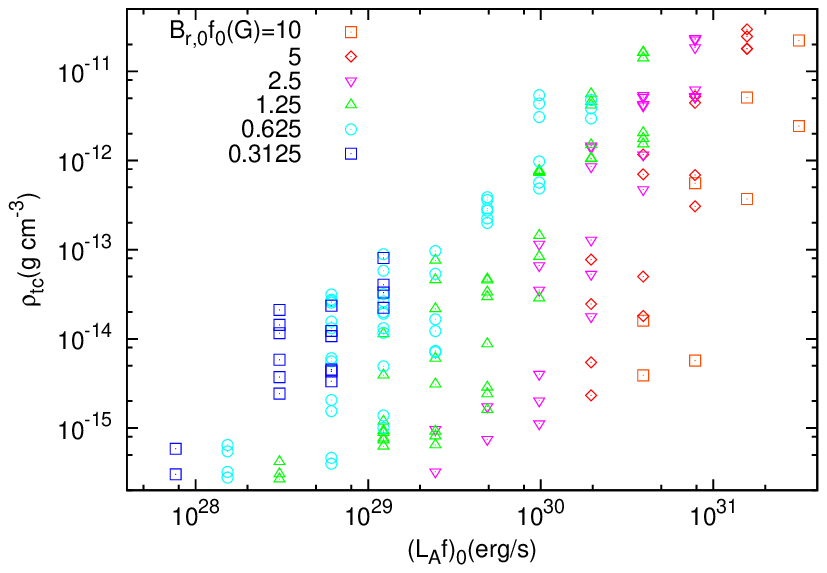}
\end{center}
\caption{{\it top}: Relation between the final kinetic energy luminosity, 
$L_{\rm K,out}$, and the input wave energy luminosity, $(L_{\rm A}f)_0$, 
at the photosphere of the simulated stellar winds. 
The lines indicate constant 
conversion factors, $c_{\rm E}=0.1,0.01,0.001$. The right and top axes 
are normalized by the solar luminosity, 
$L_{\odot}=4\times 10^{33}$ erg s$^{-1}$. 
{\it middle}: Relation between $\dot{M}$ and $(L_{\rm A}f)_0$ divided by 
$\frac{v_{\rm esc,0}^2}{2} = \frac{G M}{r_0}$. The lines indicate  
constant conversion factors, $c_{\rm M}=0.1,0.01,0.001$. 
{\it bottom}: Relation between the density, $\rho_{\rm tc}$, 
at the top of the chromosphere and $(L_{\rm A}f)_0$.}
\label{fig:Fph_SWKE}
\end{figure}

Before carrying out detailed analyses, we show the relation between 
the input \Alfven wave luminosity $(L_{\rm A}f)_0$ from the photosphere 
and the output kinetic energy, $L_{\rm K,out}$.  
The top panel of Figure \ref{fig:Fph_SWKE} shows the
$(L_{\rm A}f)_0-L_{\rm K,out}$ diagram.
In the figure we overplot three lines for constant conversion factors, 
$c_{\rm E}=0.001,0.01,0.1$, from  
$(L_{\rm A}f)_0$ to $L_{\rm K,out}$ by the three lines.  
Using Equations (\ref{eq:magflx}) and (\ref{eq:LA0}) with $v_{\rm A,0}=B_{r,0}
/\sqrt{4\pi \rho_0}$, we can explicitly write the relation that connects 
$L_{\rm K,out}$ to $(L_{\rm A}f)_0$: 
\begin{eqnarray}
L_{\rm K,out} &=& c_{\rm E} (L_{\rm A}f)_0 = c_{\rm E}\Phi_{\rm B} 
\sqrt{\frac{\rho_0}{4\pi}} \langle \delta v_0^2\rangle  \nonumber \\
&=& 2.1\times 10^{27}{\rm erg\;s^{-1}}\left(\frac{c_{\rm E}}{0.017}\right)
\nonumber \\
& & \left(\frac{B_{r,0}f_0}{1.25{\rm G}}\right)
\left\langle\left(\frac{\delta v_0}{1.34{\rm km\; s^{-1}}}\right)^2\right\rangle,
\label{eq:scke}
\end{eqnarray}
where we already substitute $r_0=R_{\odot}$ and $\rho_0=10^{-7}$ g cm$^{-3}$ 
adopted in the simulations. We discuss a more general expression for stars 
with different radius and photospheric density in \S \ref{sec:extgen}.
The relation is normalized by the input parameters and the output $L_{\rm K,oout}
=2.1\times 10^{27}$ erg s$^{-1}$ with $c_{\rm E}=0.017$ of the standard case 
for the present-day solar wind (Model Cb$0\alpha$). 
Among the four input parameters, the dependence on $B_{r,0}$, $f_0$, and 
$\delta v_0$ is explicitly shown in Equation (\ref{eq:scke}), but 
$h_{\rm l}$ does not appear because $h_{\rm l}$ causes variations of $c_{\rm E}$ 
(vertical scatters of $L_{\rm K}$ in the panel). 

Different symbols (\& colors) are used for different sets of $B_{r,0}f_0$ in 
Figure \ref{fig:Fph_SWKE}. 
We would like to note that $B_{r,0}f_0$ determines the magnetic field 
strength in the outer region where the super-radial expansion of the flux 
tubes is already completed ($f(r)\rightarrow 1$ in Equation \ref{eq:magflx}).
$B_{r,0}f_0$ is a good indicator of the properties of the solar wind, 
particularly the wind speed \citep{kfh05,suz06}.
The figure shows that the conversion factor is distributed in a range of  
$0.001<c_{\rm E}<0.1$, and a typical value is $c_{\rm E}\sim 0.01$.  

Although 
there are scatters, 
focusing on a single set of $B_{r,0}f_0$ ({\it i.e.} data points with a same 
symbol and color), $c_{\rm E}$ shows a rather clear trend which is not 
monotonic; $c_{\rm E}$ initially increases, namely $L_{\rm K,out}$ increases 
with $(L_{\rm A}f)_0$ much faster than the linear, which is followed by 
a decrease of $c_{\rm E}$ or a saturation (or even decrease) 
of $L_{\rm K,out}$ in the large $(L_{\rm A},f)_0$. 
The initial increase of $c_{\rm E}$ can be explained by the suppression 
of the reflection of the \Alfven waves \citep{suz12}, which we discuss in 
\S \ref{sec:rlrtr}.  
The later decrease is due to the enhanced radiative losses 
(\S \ref{sec:satrdls}). 
The figure shows that the saturation level of $L_{\rm K,out}$ is 
determined by $B_{r,0}f_0$, which we also discuss in \S\ref{sec:satrdls}. 

Focusing on a single set of $B_{r,0}f_0$ (a same symbol and color), 
if we input further larger $\delta v_0$ to increase $(L_{\rm A}f)_0$, 
$L_{\rm k,out}$ will be smaller to give $c_{\rm E}<0.001$ and to fill the 
lower right corner of the panel. 
However, the largest $\delta v_0 =7.57$ km s$^{-1}$ is 
already supersonic driving at the photosphere 
(note $c_{\rm s,0}= 6.31$ km s$^{-1}$), which is probably extremely large. 
Therefore, we can conclude that 
the conversion factor is located well in the range of $0.001<c_{\rm E}<0.1$ 
in the realistic situations. 

The top panel of Figure \ref{fig:Fph_SWKE} also shows vertical scatters. 
This indicates that the final wind kinetic energies are different 
even though the input $(L_{\rm A}f_0)$$(\propto B_{r,0}f_0 \delta v_0^2)$'s are 
identical. This is mainly because of the difference of $h_{\rm l}$, which 
controls the reflection and transmission of the \Alfven waves in the 
chromosphere, which we will discuss in \S \ref{sec:rlrtr}.

It is more useful for readers to show the mass loss rates as a function of 
surface properties. 
The middle panel of Figure \ref{fig:Fph_SWKE} shows $\dot{M}$ 
with wave luminosity divided by $\frac{v_{\rm esc,0}^2}{2}=\frac{GM}{r_0}$ 
in order that both axes are in unit of $M_{\odot}$yr$^{-1}$, where 
$v_{\rm esc,0}$($=618$ km s$^{-1}$) is the escape velocity. 
Similarly to the scaling relation for $L_{\rm K,out}$ (Equation \ref{eq:scke}), 
we can define a conversion factor, $c_{\rm M}$, with respect to the 
mass loss rates, and then, we have 
\begin{eqnarray}
\dot{M} &=& c_{\rm M}\frac{(L_{\rm A}f)_0 r_0}{GM}
=c_{\rm M}\Phi_{\rm B} \sqrt{\frac{\rho_0}{4\pi}}
\frac{\langle \delta v_0^2\rangle r_0}{GM}, \nonumber \\
&=& 2.2\times 10^{-14}M_{\odot}{\rm yr}^{-1}\left(\frac{c_{\rm M}}{0.023}\right)
\nonumber \\
& &
\left(\frac{B_{r,0}f_0}{1.25{\rm G}}\right)
\left\langle\left(\frac{\delta v_0}{1.34{\rm km\; s^{-1}}}\right)^2\right\rangle
\label{eq:mssc}
\end{eqnarray}
where we already substitute $M=M_{\odot}$, $r_0=R_{\odot}$, and $\rho_0=10^{-7}$
g cm$^{-3}$ that are adopted in the simulations, but see \S \ref{sec:extgen} 
for a more general expression. 
Here, the normalizations are again adopted from the input parameters 
and the results ($\dot{M}=2.2\times 10^{-14}M_{\odot}$yr$^{-1}$ and 
$c_{\rm M}=0.023$) of the standard case for the present-day Sun.  

In the bottom panel of Figure \ref{fig:Fph_SWKE}, we plot the 
density, $\rho_{\rm tc}$, at the top of the chromosphere defined at 
$T=2\times 10^4$ K.  
$\rho_{\rm tc}$, which essentially corresponds to the so-called base density 
for the stellar winds, is determined by the energy balance among the
wave heating in the above corona, the downward thermal conduction, and 
the radiative cooling in the transition region as described in 
\S \ref{sec:engfm}. 
A larger energy injection, $(L_{\rm A}f)_0$, from the photosphere 
leads to larger heating in the corona. Then, the downward thermal conduction 
from the corona is enhanced, which increases $\rho_{\rm tc}$ by the 
chromospheric evaporation.  
Comparison of the bottom panel to the top and middle panels show that 
the differences of $L_{\rm K,out}$ and $\dot{M}$ that extent more than orders of 
magnitude among different cases can be mainly explained by the differences 
of $\rho_{\rm tc}$. 
Moreover, the scatters of $L_{\rm K,out}$ on a fixed 
$(L_{\rm A}f)_0$ are smaller than the scatters of $\rho_{\rm tc}$. 
This is mainly because the wind speed becomes slower as 
the wind density ($\propto \rho_{\rm tc}$) increases with increasing 
$(L_{\rm A}f)_0$; the dense wind cannot be accelerated to high speed. 
Related to this, the top panel of Figure \ref{fig:Fph_SWKE} illustrates that 
the average trend of $c_{\rm E}$ does not show a systematic trend with 
$(L_{\rm A}f)_0$, while the middle panel still shows that the average trend 
of $c_{\rm M}$ is slightly increasing with $\frac{(L_{\rm A}f)_0 r_0}{G M}$. 
However, this issue needs a caution.
If a star rotates very rapidly, the wind speed of active cases 
(on the right side of the panels) might be higher. 
In this case $c_{\rm E}$ might show an increasing trend with 
$(L_{\rm A}f)_0$, similarly to $c_{\rm M}$. 

\subsection{Initial Increase of $L_{\rm K,out}$ --Wave Reflection--}
\label{sec:rlrtr}
\subsubsection{Energetics}
\begin{figure}
\begin{center}
\FigureFile(90mm,){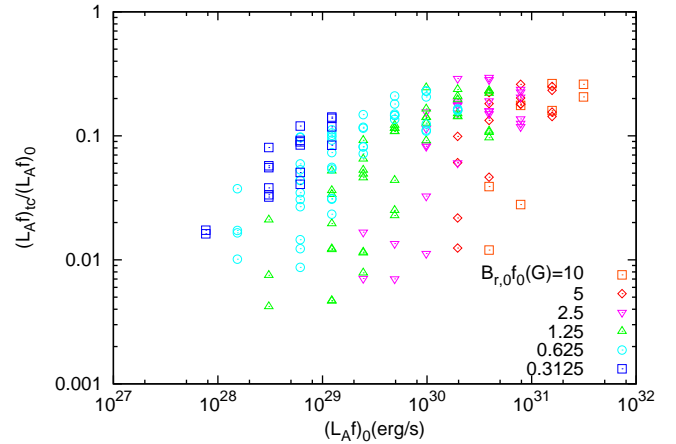}
\end{center}
\caption{Surviving fraction, $c_{\rm T}=(L_{\rm A}f)_{\rm tc}/(L_{\rm A}f)_0$, 
of the \Alfven wave luminosity at the top of the chromosphere versus  
input wave luminosity at the photosphere, $(L_{\rm A}f)_0$. }
\label{fig:Fph_Ftr}
\end{figure}

\begin{table*}
{\footnotesize
\begin{center}
\begin{tabular}{|c|c|c|c|c||c|c|c|c|c|c|}
\hline
Model & $\delta v_0$  & $B_{r,0}$(kG)$\times f_0$ & 
$h_{\rm l}$ ($r_0$) & $(L_{\rm A}f)_0$ & $(L_{\rm A}f)_{\rm tc}$ 
& $c_{\rm T}$
& $\dot{M}$ ($M_{\odot}$yr$^{-1}$) & $v_{r,{\rm out}}$ 
& $L_{\rm K,out}$ & $(L_{\rm R}f)_{\rm tc}$  \\
\hline
\hline
Cb$0\alpha$ & 1.34 & ${1}/{800}$ & 0.01 & $1.2\times 10^{29}$ 
& $6.4\times 10^{27}$ & 0.052 & $2.2\times 10^{-14}$ & $554$ 
& $2.1\times 10^{27}$ & $1.5\times 10^{27}$  \\
\hline
\hline
Cc$-2\beta$ & 0.669 & ${2}/{1600}$ & 0.03 & $3.1\times 10^{28}$ 
& $1.3\times 10^{26}$ & 0.0042 & $1.5\times 10^{-16}$ & $932$ 
& $4.2\times 10^{25}$ & $1.2\times 10^{25}$  \\
\hline
Cc$0\beta$ & 1.34 & ${2}/{1600}$ & 0.03 & $1.2\times 10^{29}$ 
& $1.5\times 10^{27}$ & 0.012 & $2.5\times 10^{-15}$ & $933$ 
& $7.0\times 10^{26}$ & $4.8\times 10^{25}$  \\
\hline
Cc$+2\beta$ & 2.68 & ${2}/{1600}$ & 0.03 & $4.9\times 10^{29}$ 
& $5.7\times 10^{28}$ & 0.12 & $2.5\times 10^{-13}$ & $438$ 
& $1.5\times 10^{28}$ & $9.2\times 10^{27}$  \\
\hline
Cc$+4\beta$ & 5.35 & ${2}/{1600}$ & 0.03 & $2.0\times 10^{30}$ 
& $3.0\times 10^{29}$ & 0.15 & $1.1\times 10^{-12}$ & $239$ 
& $1.9\times 10^{28}$ & $3.8\times 10^{29}$  \\
\hline
\end{tabular}
\end{center}
}
\caption{Input parameters and outputs for the cases shown in Figures 
\ref{fig:tmavstr1} \& \ref{fig:tmavwav1}. The unit of $\delta v_0$ 
and $v_{r,{\rm out}}$ is km s$^{-1}$, and the unit of $L$ is erg s$^{-1}$. 
$c_{\rm T}$ is the transmission fraction of the \Alfven waves from the 
photosphere to the top of the chromosphere (Equation \ref{eq:trnsm}).
Model Cb$0\alpha$ is the reference case for the present sun, 
and the lower four Models are the cases with different $\delta v_0$ 
by twice each but with the same $B_{r,0}f_0$ and $h_{\rm l}$. }
\label{tab:ref}
\end{table*}

A sizable fraction of the input \Alfven waves suffers reflection before 
reaching the corona \citep{hol84,moo91,mm10,ver12}. Observation by HINODE 
obtained signatures of reflected \Alfven waves at the photosphere 
\citep{ft09}. In the chromosphere, 
the density decreases very rapidly, which leads to the rapid increase 
of the \Alfven velocity. Therefore, the shape of outwardly propagating 
\Alfven waves is considerably deformed, which indicates that these
waves are reflected back downward. 
For the current solar condition, typically $\gtrsim$ 90 \% of the input 
energy is reflected back downward \citep{szi06,ms12}.

We measure $(L_{\rm A}f)_{\rm tc}$ at the top of the chromospheres, 
$r=r_{\rm tc}$, to quantitatively study the surviving fractions 
($=c_{\rm T}$ in Equation \ref{eq:trnsm}) after the propagation through 
the chromospheres. 
The reflected waves are mostly dissipate through the downward propagation 
in the chromospheres and finally lost by the radiation, where some fractions 
are considered to be again reflected upward and may be trapped 
\citep{ms10}. 
Therefore, the radiation loss in the chromospheres is mostly taken into account 
in the component of the reflected waves. 

Figure \ref{fig:Fph_Ftr} displays 
$c_{\rm T}=(L_{\rm A}f)_{\rm tc}/(L_{\rm A}f)_0$.
For small $(L_{\rm A}f)_0$, the reflection 
of the \Alfven waves is so effective that only $\sim 1$\% of the input energies 
can transit through the chromospheres.  On the other hand, the fraction is 
increasing with increasing $(L_{\rm A}f)_0$ and finally reaches $\sim$ 0.3. 
This trend can be understood from the density structures in the chromospheres, 
which is discussed later (Fig.\ref{fig:tmavstr1}). 
In short, in cases with large energy inputs the reflection of the outgoing 
\Alfven waves is not so effective because the densities decrease more slowly 
and accordingly the change of the \Alfven speeds is more gradual.

The suppression of the reflection of the \Alfven waves (the increase of 
$c_{\rm T}$ with $(L_{\rm A}f)_0$ 
) is the main reason of the initial rise of 
the energy conversion factor, $c_{\rm E}$($=L_{\rm K,out}/(L_{\rm A}f)_0$;  
Equation \ref{eq:scke}) shown in Fig.\ref{fig:Fph_SWKE}. 
A larger fraction of the \Alfven waves transmits to 
upper regions, which directly leads to the larger final kinetic energy 
of the stellar wind.  

Figure \ref{fig:Fph_Ftr} also shows that $c_{\rm T}$ largely 
scatters even for the same sets of $\delta v_0$ and $B_{r,0}f_0$ (same symbol 
and color at same $(L_{\rm A}f)_0$) because of the effect of different 
$h_{\rm l}$. Larger 
$h_{\rm l}$ indicates that the magnetic field strength in the chromospheric 
region is larger, and the \Alfven speed more rapidly increases 
due to the decrease of the density. As a result, a larger fraction of 
the \Alfven waves from the photosphere is reflected back, giving smaller 
$c_{\rm T}$. 

\subsubsection{Atmospheric Structure}

\label{sec:refstr}

\begin{figure}
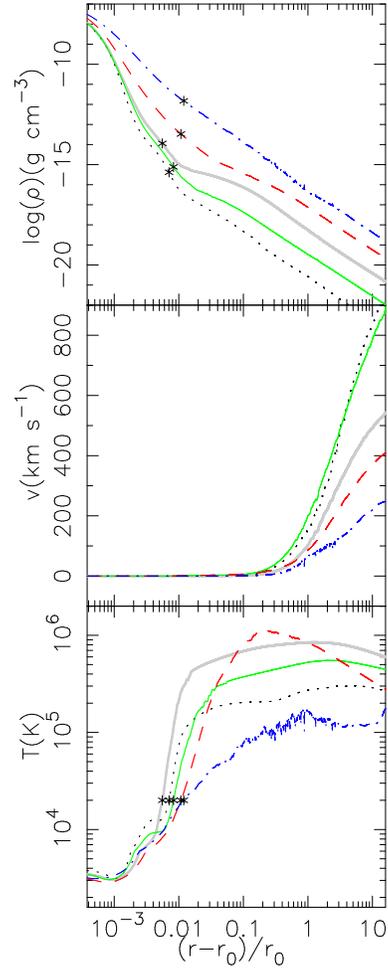

\begin{center}
\FigureFile(50mm,){Ysunstrave_cmp3.ps}
\end{center}
\caption{Comparison of the time-averaged wind structures in 
Table \ref{tab:ref}. The four cases with the same 
$B_{r,0}f_0=2({\rm \; kG})/1600$ and $h_{\rm l}=0.03 r_0$ but different 
$\delta v_0$ by twice each (Cc$-2\beta$: black dotted, Cc$0\beta$: 
green solid, Cc$+2\beta$: red dashed, and Cc$+4\beta$: blue dot-dashed) 
are compared with the reference case for the present sun 
(Cb0$\alpha$: thick gray solid).
From top to bottom, the densities, radial velocities, and temperatures are
compared. In the top and bottom panels, the location of the 
top of the chromosphere at $T=2\times 10^4$ K of each case is shown 
by asterisks.} 
\label{fig:tmavstr1}
\end{figure}

We demonstrate simulated wind structures with different energy inputs. 
In Figure \ref{fig:tmavstr1}, we plot the time-averaged structures of 
the four cases with the same $B_{r,0}f_0=2({\rm \; kG})/1600$ 
and $h_{\rm l}=0.03 r_0$ but different $\delta v_0$ by twice each 
in comparison with the case for 
the present sun (Table \ref{tab:ref}). The input energy of the four cases 
is different by four times each since $(L_{\rm A}f)_0\propto B_{r,0} f_0 
\delta v_0^2$.

\begin{figure}
\begin{center}
\FigureFile(90mm,){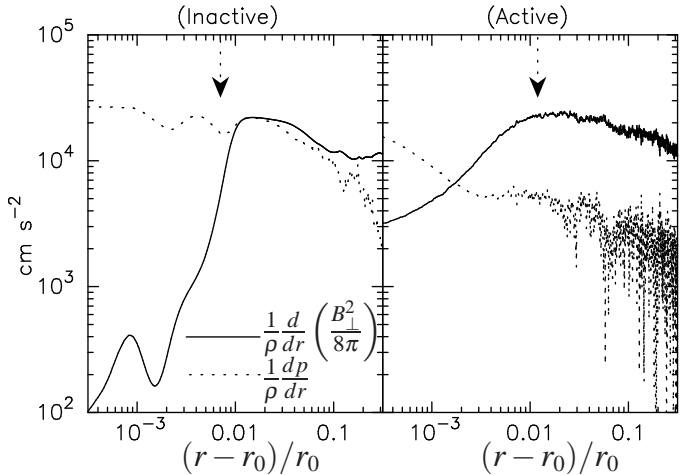}
\end{center}
\caption{Comparison of the accelerations (forces per mass) by the magnetic 
pressure gradient 
(solid lines) and the gas pressure gradient (dashed lines) of the most  
inactive case (Model Cc$-2\beta$; left) and the most active case 
(Model Cc$+4\beta$; right) in Table \ref{tab:ref}. The location 
of the top of the chromosphere at $T=2\times 10^4$ K is also shown by the 
vertical arrow in each panel.}
\label{fig:chdens}
\end{figure}

The top panel of Figure \ref{fig:tmavstr1} indicates that the decrease of the 
density becomes slower with increasing the energy input from the surface. 
This is mainly because in the active cases the magnetic pressure 
dominantly supports the chromospheric density structure in addition to
the gas pressure. 
Figure \ref{fig:chdens} compares the force (per mass) by the magnetic pressure 
gradient (solid lines) with the force by the gas pressure gradient (dashed 
lines) of the most inactive case (Model Cc$-2\beta$; left panel, 
corresponding to the black dotted lines in Figure \ref{fig:tmavstr1})
and the most active case (Model Cc$+4\beta$; right panel, 
corresponding to the blue 
dot-dashed lines in Figure \ref{fig:tmavstr1}) in Table \ref{tab:ref}. 
The left panel shows that in the inactive case the gas pressure gradient 
dominates in the chromosphere and the transition region, 
$r-r_0<10^{-2}r_0$.
On the other hand, in the active case (right panel)
the magnetic pressure gradient dominantly supports the density structure at 
and above the middle chromosphere, $r-r_0> 2\times 10^{-3}r_0$, because the 
input wave amplitude ($\delta v_{\perp} \propto \delta B_{\perp}$) from the 
surface itself is larger. 
As a result, the cool chromospheric material extends to an upper region
The density of the upper region is also larger, which leads to the larger 
$\dot{M}$.

\begin{figure}
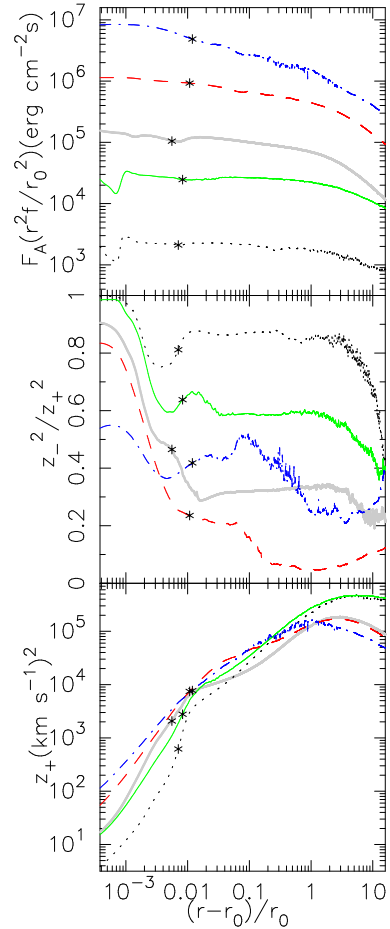

\begin{center}
\FigureFile(50mm,){Ysunwaveave_cmp2.ps}
\end{center}
\caption{Comparison of time-average \Alfven wave properties of the three cases 
tabulated in Table \ref{tab:ref}. The line types are the same as in 
Figure \ref{fig:tmavstr1}.  The reference case for the current 
sun is shown (Model Cb0$\alpha$; thick gray solid) in comparison, and the 
other four cases adopt the different input Poynting flux, $(L_{\rm A}f)_0$, 
from the photosphere by 4 times each.  The location of the 
top of the chromosphere at $T=2\times 10^4$ K of each case is shown 
by asterisks.
{\it top}: Energy flux of the \Alfven waves. 
In order to remove the effect of the expansion of the flux tubes, we 
display $F_{\rm A}r^2 f/r_0^2 = L_{\rm A}f /4\pi r_0^2$ 
(see Equation \ref{eq:Alfflx} or \ref{eq:Alflum}), {\it middle}: Ratio 
of the outgoing and incoming (reflected) components of the \Alfven waves, 
$z_{-}^2/z_{\rm +}^2$. {\it Bottom}: Squared Els\"{a}sser variable, $z_{+}^2$. }
\label{fig:tmavwav1}
\end{figure}

The change of the density structure directly affects the reflection of 
the \Alfven waves. 
In the chromosphere the \Alfven speed, $v_{\rm A}=B_r/\sqrt{4\pi\rho}$, 
generally increases with height because of the decrease 
of the density. The slower decrease of the density in the active case 
makes the change of $v_{\rm A}$ more gradual. 
Then, the reflection of the \Alfven waves is not so severe in the active 
case, which we will discuss in \S \ref{sec:wvprp} and 
Figure \ref{fig:tmavwav1}.
Table \ref{tab:ref} clearly indicates that the transmissivity 
($c_{\rm T}=(L_{\rm A}f)_{\rm tc}/(L_{\rm A}f)_0$) of the \Alfven waves through 
the chromosphere increases with the input energy as shown in 
Figure \ref{fig:Fph_Ftr}. Accordingly, 
the kinetic energy luminosity, $L_{\rm K,out}$, and the mass loss rate, 
$\dot{M}$, rapidly increase with $(L_{\rm A}f)_0$. 
The radiation loss, $(L_{\rm R}f)_{\rm tc}$, also rapidly increases with 
$(L_{\rm A}f)_0$, and its dependence is faster than the dependence of 
$L_{\rm K,out}$, 
which we discuss in \S \ref{sec:satrdls} in terms of the saturation of 
the stellar winds. 

The middle panel of Figure \ref{fig:tmavstr1} shows that the wind velocity 
becomes slower with increasing the energy input because 
more mass is lifted up to the wind region, 
and then, the wind material cannot be effectively accelerated to higher 
velocities. 

The bottom panel of Figure \ref{fig:tmavstr1} indicates that the maximum 
temperature increases with increasing the input 
energy up to the second most active case with $\delta v_0= 2.68$ km s$^{-1}$
(Model Cc$+2\beta$; red dashed). On the other hand, in the most active case 
with $\delta v_0= 5.35$ km s$^{-1}$ (Model Cc$+4\beta$; blue dot-dashed) the 
temperature is low without the steady hot corona with $T\gtrsim 10^6$ K, 
because the radiative cooling ($\propto \rho^2$ in the optically thin 
plasma) is efficient in the dense circumstances.  As a result, the coronal 
temperature cannot be maintained, and the temperature time-dependently goes 
up and down from $\sim 10^4$ K to $\gtrsim 10^6$ K (\S \ref{sec:evchr}) 
because of the thermal instability in the radiative cooling function 
\citep{sd93,suz07}.
The plotted value of temperature, $\sim 10^{5}$ K, is the time-averaged value.

The temperature of the second most active case still drops faster than 
the cases with smaller $\delta v_0$ because the thermal conduction is 
relatively ineffective in comparison with the adiabatic cooling in 
the high density circumstances.   

\label{sec:rgrtr}
\begin{table*}
{\footnotesize 
\begin{tabular}{|c|c|c|c|c||c|c|c|c|c|c|}
\hline
Model & $\delta v_0$ & $B_{r,0}$(kG)$\times f_0$ &
$h_{\rm l}$ ($r_0$) & $(L_{\rm A}f)_0$ & $(L_{\rm A}f)_{\rm tc}$ 
& $c_{\rm T}$ 
& $\dot{M}$ ($M_{\odot}$yr$^{-1}$) & $v_{r,{\rm out}}$ 
& $L_{\rm K,out}$ & $(L_{\rm R}f)_{\rm tc}$ \\
\hline
\hline
Ac$+2\gamma$ &  2.68 & ${2}/{6400}$ & 0.1 & $1.2\times 10^{29}$ 
& $1.7\times 10^{28}$ & 0.14 & $1.1\times 10^{-13}$ & $226$
& $1.8\times 10^{27}$ & $2.4\times 10^{27}$\\
\hline
Bc$+4\beta$ & 5.35 & ${2}/{3200}$ & 0.03 & $9.8\times 10^{29}$ 
& $1.1\times 10^{29}$ & 0.11 & $5.6\times 10^{-13}$ & $278$
& $1.4\times 10^{28}$ & $1.8\times 10^{29}$ \\
\hline
Cd$+3\gamma$ & 3.79 & ${4}/{3200}$ & 0.1 & $9.8\times 10^{29}$ 
& $1.6\times 10^{29}$ & 0.17 & $8.5\times 10^{-13}$ & $313$
& $2.6\times 10^{28}$ & $2.2\times 10^{28}$\\
\hline
Df$+3\gamma$ & 3.79 & ${16}/{6400}$ & 0.1 & $2.0\times 10^{30}$ 
& $5.7\times 10^{29}$ & 0.29 & $2.5\times 10^{-12}$ & $373$
& $1.1\times 10^{29}$ & $7.2\times 10^{28}$\\
\hline
Ef$+4\gamma$ & 5.35 & ${16}/{3200}$ & 0.1 & $7.8\times 10^{30}$ 
& $2.1\times 10^{30}$ & 0.26 & $1.0\times 10^{-11}$ & $351$
& $3.9\times 10^{29}$ & $4.5\times 10^{29}$\\
\hline
Ff$+5\gamma$  & 7.57 & ${16}/{1600}$ & 0.1 & $3.1\times 10^{31}$ 
& $8.2\times 10^{30}$ & 0.26 & $4.3\times 10^{-11}$ & $317$
& $1.4\times 10^{30}$ & $2.9\times 10^{30}$\\
\hline
\end{tabular}
}
\caption{Saturated cases for $B_{r,0}f_0=0.3125,0.625,1.25,2.5,5,10$(G) from 
top to bottom, shown in Figure \ref{fig:sat1}. 
The unit of $\delta v_0$ and $v_{r,{\rm out}}$ is 
km s$^{-1}$, and the unit of $L$ is erg s$^{-1}$. $c_{\rm T}$ is the surviving 
fraction of the \Alfven waves from the photosphere to the top of the 
chromosphere (Equation \ref{eq:trnsm}).
\label{tab:sat2}
}
\end{table*}


\subsubsection{Wave Properties}
\label{sec:wvprp}

Figure \ref{fig:tmavwav1} displays properties of the \Alfven waves 
of the same cases shown in Figure \ref{fig:tmavstr1} and Table 
\ref{tab:ref}. 
The top panel of Figure \ref{fig:tmavwav1} compares the \Alfven wave 
energy fluxes (Equation \ref{eq:Alfflx}). 
Here we plot $F_{\rm A} r^2 f/r_0^2 = L_{\rm A} f/4\pi r_0^2$ to exclude 
the effect of the adiabatic expansion in the super-radially open flux tubes. 
While the input Poynting flux ($(L_{\rm A}f)_0\propto B_0 f_0 \delta v_0^2$) 
from the photosphere is different by 4 times each,
the differences of the wave energy fluxes in the 
atmospheres are more than that extent. 
In particular, the difference is more than one order of magnitude 
among the three cases, Model Cc$-2\beta$ (black dotted), Model 
(Cc$0\beta$; green solid), and Model Cc$+2\beta$ (red dashed). 
The sensitive dependence of the wave energy flux in 
the atmosphere on the input wave energy is because of the suppression of the
reflection of the outgoing \Alfven waves 
as discussed so far. 

The middle panel of Figure \ref{fig:tmavwav1}, which compares 
$z_{-}^2/z_{+}^2$, the ratio of the reflected waves to the outgoing waves, 
exhibits that the wave reflection is suppressed for larger 
$(L_{\rm A}f)_0$. One exception is the difference between 
the second most active case (Cc$+2\beta$; red dashed) and the most active 
case (Cc$+4\beta$; blue dot-dashed). In the upper region, $r-r_0 > 0.01 r_0$, 
the fraction of the reflected component is larger in the most active case, 
which is opposed to the tendency for the cases with the smaller energy inputs. 
This is because in the most active case the density decreases faster 
in the upper region as a result of the smaller (time-averaged) temperature. 
Then, a finite level of the reflection continues even in the upper region, which 
also results in the faster decrease of $F_{\rm A}$ (upper panel of Figure 
\ref{fig:tmavwav1}). 

The bottom panel of Figure \ref{fig:tmavwav1} compares $z_{+}^2$. This 
interestingly shows that the wave amplitudes of the outgoing component 
are comparable in these cases although the energy fluxes are very 
different. 
The difference of $F_{\rm A}$ is mostly owing to the difference of the 
densities, in addition to the difference of the reflection efficiency.

\subsection{Saturation of $L_{\rm K,out}$ --Enhanced Radiative Loss--}
\label{sec:satrdls}
\subsubsection{Energetics}

\begin{figure}
\begin{center}
\FigureFile(90mm,){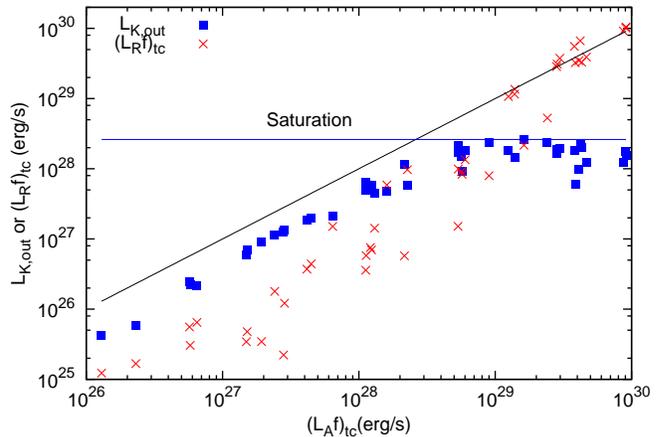}
\end{center}
\caption{Kinetic energy, $L_{\rm K,out}$, (filled squares) and radiation loss, 
$(L_{\rm R}f)_{\rm tc}$ (crosses) versus upgoing Poynting flux, 
$(L_{\rm A}f)_{\rm tc}$, at the top of the chromosphere of the cases with 
$B_{r,0}f_0=1.25$ G. The diagonal line indicates the $x=y$ relation and 
the horizontal line indicates the saturation level of $L_{\rm K,out}$.}
\label{fig:Ftr_dep-Bf2}
\end{figure}

\begin{figure}
\begin{center}
\FigureFile(90mm,){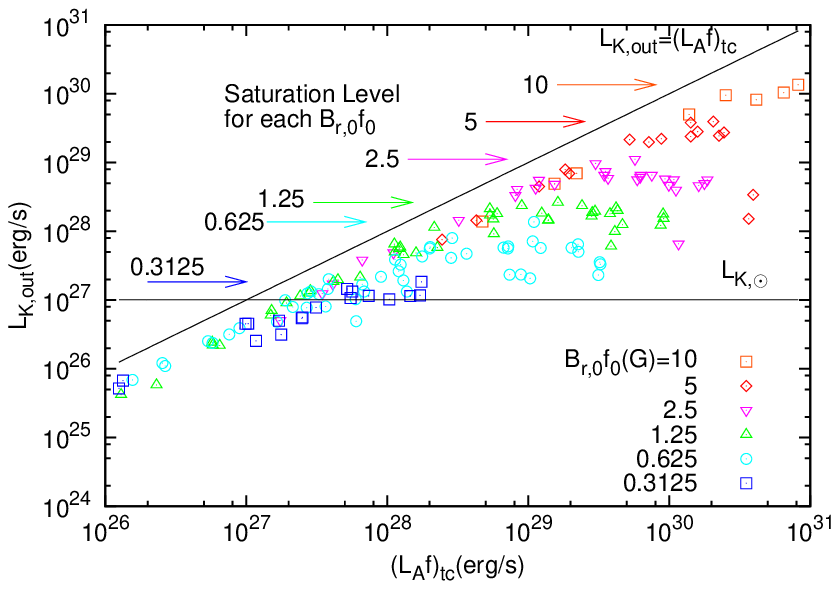}
\FigureFile(90mm,){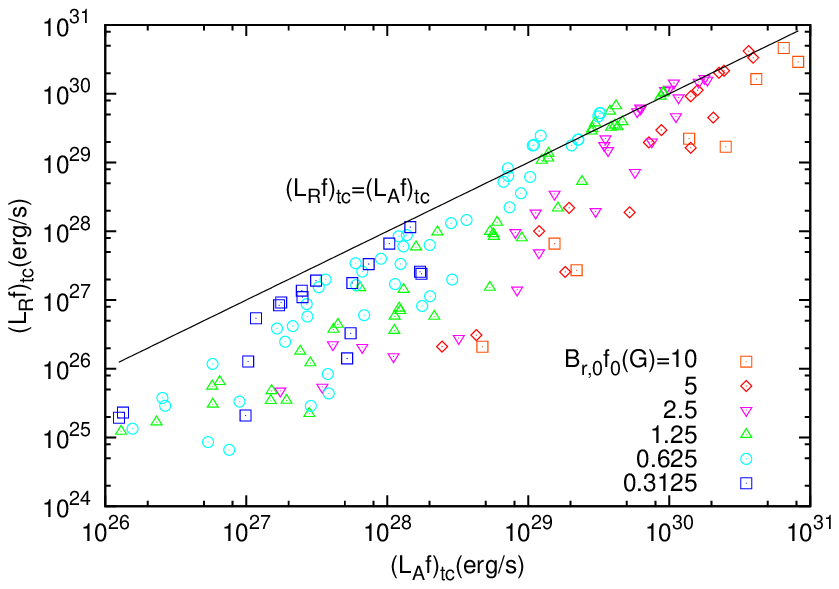}
\FigureFile(90mm,){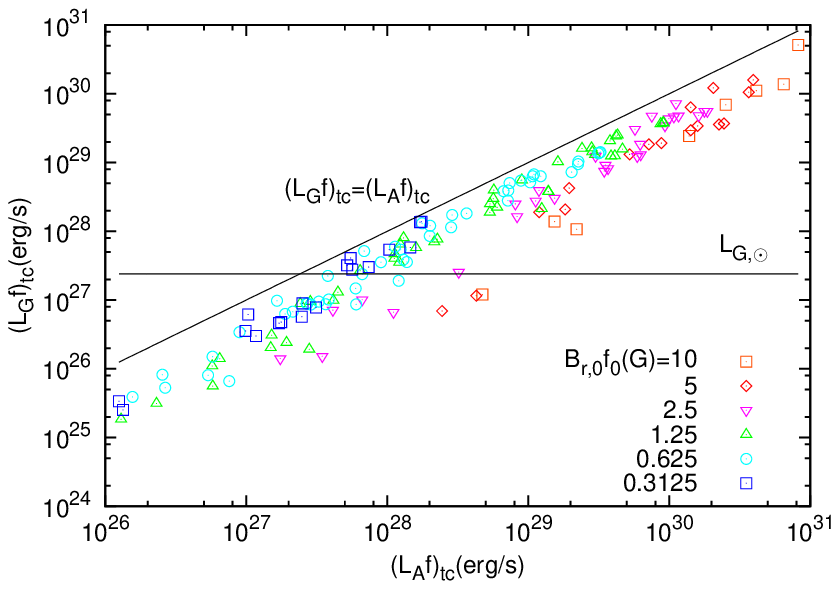}
\end{center}
\caption{Energetics of the simulated stellar winds. From top to bottom, 
kinetic energy, $L_{\rm K,out}$, radiation loss, $(L_{\rm R}f)_{\rm tc}$, 
and gravitational loss, $(L_{\rm G}f)_{\rm tc}$, are shown on 
upgoing Poynting flux, $(L_{\rm A}f)_{\rm tc}$, at the top of the chromosphere.
The diagonal line in each panel indicates the $x=y$ relation. In the 
panels for $L_{\rm K,out}$ and $(L_{\rm G}f)_{\rm tc}$, the present 
solar level which is estimated by using $\dot{M}=2\times 10^{-14}$ 
$M_{\odot}$yr$^{-1}$ and $v_r=400$ km s$^{-1}$ is also shown by the horizontal 
lines. 
In the top panel, the saturation level of $L_{\rm K,out}$ for 
each $B_{r,0}f_0$ is indicated by the color arrows. 
} 
\label{fig:Ftr_dep}
\end{figure}

We examine the energetics of the stellar winds above the chromosphere 
by Equation (\ref{eq:engsw1}).
In Figure \ref{fig:Ftr_dep-Bf2}, we show the wind kinetic energy luminosities, 
$L_{\rm K,out}$ (filled squares), 
and the radiation losses, $(L_{\rm R}f)_{\rm tc}$ (crosses), with the net 
outgoing Poynting fluxes, $(L_{\rm A}f)_{\rm tc}$, measured at the top 
of the chromosphere where $T=2\times 10^4$ K of the cases with 
$B_{r,0}f_0=1.25$ G. 
The fraction of the wind kinetic 
energy, $L_{\rm K,out}/(L_{\rm A}f)_{\rm tc}$, is initially large, 
with roughly $\sim$ half of $(L_{\rm A}f)_{\rm tc}$ transferred 
to the stellar wind.  $L_{\rm K,out}$ increases almost linearly with 
$(L_{\rm A}f)_{\rm tc}$,  but eventually saturates 
for large $(L_{\rm A}f)_{\rm tc}$. 
Some cases even show considerably smaller $L_{\rm K,out}$ than other cases 
with similar $(L_{\rm A}f)_{\rm tc}$. 
The saturation of the wind kinetic energy is a consequence of the enhancement 
of the radiation loss. 
$(L_{\rm R}f)_{\rm tc}$ increases with 
$(L_{\rm A}f)_{\rm tc}$ much faster than the linear dependence, because the wind 
density increases as a result of strong driving as we increase the input 
energy. In the optically thin plasma, the radiation loss is proportional 
to $\rho^2$, and then, $(L_{\rm R}f)_{\rm tc}$ rapidly increases with 
$(L_{\rm A}f)_{\rm tc}$. 
The fraction of the radiation loss, 
$(L_{\rm R}f)_{\rm tc}/(L_{\rm A}f)_{\rm tc}$,  which starts from a much 
smaller value at small $(L_{\rm A}f)_{\rm tc}$ than 
$L_{\rm K,out}/(L_{\rm A}f)_{\rm tc}$, 
is approaching to $\approx$ unity\footnote{In some cases, the radiative losses 
exceed the injected Poynting fluxes from the top of the chromosphere, 
$(L_{\rm R}f)_{\rm tc}/(L_{\rm A}f)_{\rm tc}>1$, which seems to break 
the energy conservation relation. 
In addition to the Poynting fluxes, however, there are contributions 
from perturbations of gas pressure (Eq.\ref{eq:engeq2}) to the energy 
injections, which we do not take into account in
Equation (\ref{eq:engsw1}). The gas pressure perturbations are due to 
sound waves nonlinearly generated from \Alfven waves \citep{ks99,szi06}. 
Although the energy inputs from such sound waves are smaller than 
the energy inputs from the \Alfven waves, they are not negligible in 
active cases. }
($(L_{\rm R}f)_{\rm tc}=(L_{\rm A}f)_{\rm tc}$) for large 
$(L_{\rm A}f)_{\rm tc}$. In this saturated state, the Poynting flux 
passing through the top of the chromosphere is mostly radiated away 
in the upper regions.
As a result, the fractions transferred to the wind kinetic 
energies are quite small in these cases, which is observed as the saturation 
or drop of $L_{\rm K,out}$ with increasing $(L_{\rm A}f)_{\rm tc}$.

Figure \ref{fig:Ftr_dep} displays each term of Equation (\ref{eq:engsw1}), 
the kinetic energy luminosity, $L_{\rm K,out}$ (top panel), the radiation loss, 
$(L_{\rm R}f)_{\rm tc}$ (middle panel), and the gravitational loss, 
$(L_{\rm G}f)_{\rm tc}$ (bottom panel), of the stellar winds 
with the net outgoing Poynting flux, $(L_{\rm A}f)_{\rm tc}$, 
measured at the top of the chromosphere. 
Figure \ref{fig:Ftr_dep} clearly shows that the trend --the saturation of 
$L_{\rm K,out}$ as a consequence of the enhanced $(L_{\rm R}f)_{\rm tc}$ --
is universal for the other values of $B_{r,0}f_0$. Furthermore, 
the saturation level of the kinetic energy luminosity, $L_{\rm K,out,sat}$, 
is positively correlated with 
$B_{r,0}f_0$ as shown in the top panel, where $L_{\rm K,out,sat}$ is 
taken from the case which gives the largest $L_{\rm K,out}$ for each 
$B_{r,0}f_0$ (Table \ref{tab:sat2}).  
The nonlinear dissipation of the \Alfven waves can explain the dependence 
of the saturation level on $B_{r,0}f_0$. 
Since $B_r=B_{r,0}f_0r_0^2/r^2$, $B_{r,0}f_0$ determines the
\Alfven speed, $v_{\rm A}$, in the wind acceleration region 
($1.5r_0\lesssim r \lesssim 10r_0)$). 
The nonlinearity of \Alfven waves, $\delta v_\perp/v_{\rm A}$, is an important 
factor which controls the dissipation of \Alfven waves, and 
larger $\delta v_\perp/v_{\rm A}$ leads to faster dissipation 
(e.g. Hollweg 1973; Suzuki \& Inutsuka 2006).
Then, in cases with smaller $B_{r,0}f_0$, the nonlinearity becomes relatively 
larger because of the smaller $v_{\rm A}$ in the wind acceleration regions 
\citep{suz12}.  
In these cases, the \Alfven waves dissipate faster at lower altitudes where 
the densities are higher. As a result, more wave energy is transferred to 
the radiation losses rather than the wind kinetic energy. 
On the other hand, in cases with larger $B_{r,0}f_0$ the \Alfven waves 
can travel to higher altitudes and effectively drive the stellar winds, which 
gives the higher saturation levels of $L_{\rm K,out}$. 
Although our simulations focus on the nonlinear generation of 
compressive waves as the primary dissipation mechanism of the \Alfven 
waves, the discussion on the saturation so far can be applied provided that 
nonlinear dissipation processes, e.g. turbulent cascade, dominantly operate 
because $\delta v_{\perp}/v_{\rm A}$ controls the wave dissipation in the same 
manner. 
On the other hand, if the dissipation of linear \Alfven waves, e.g. phase 
mixing due to viscosity and resistivity \citep{hp83} or Landau-type damping 
due to kinetic effects (e.g. Suzuki et al.2006), was dominant, the saturation 
would give a different tendency. 

\begin{figure}
\begin{center}
\FigureFile(80mm,){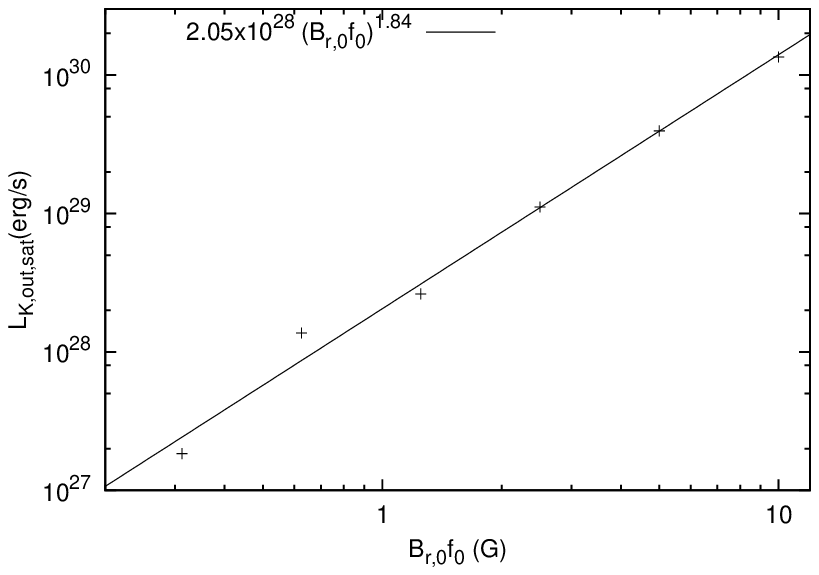}
\FigureFile(80mm,){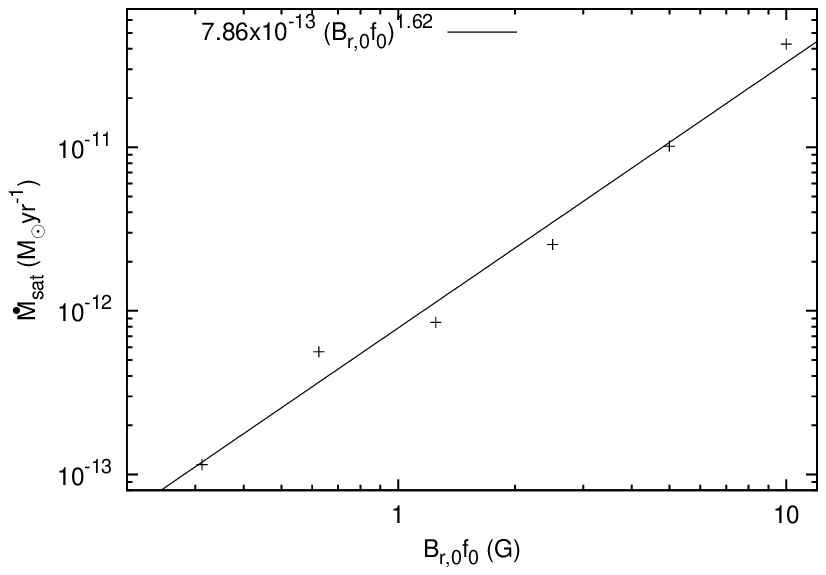}
\end{center}
\caption{{\it Upper panel}: Kinetic energy, $L_{\rm K,out,sat}$, of the 
saturated cases for the respective $B_{r,0}f_0$ with $B_{r,0}f_0$. 
The line is the best fit function, Equation (\ref{eq:fitke}). 
{\it Lower panel}: Mass loss rate, $\dot{M}_{\rm sat}$ of the saturated 
cases. The line is the best fit function, Equation (\ref{eq:fitmdot}).
}
\label{fig:sat1}
\end{figure}

We show the wind kinetic energy, 
$L_{\rm K,out,sat}$, and the corresponding mass loss 
rate, $\dot{M}_{\rm sat}$, of the saturated cases (Table \ref{tab:sat2}) 
in Figure \ref{fig:sat1}.  
Both $L_{\rm K,out,sat}$ and $\dot{M}_{\rm sat}$ are well fitted by power-law 
functions (solid lines in Figure \ref{fig:sat1}),
\begin{equation}
L_{\rm K,out,sat} = 2.05\times 10^{28}{\rm erg\;s^{-1}}(B_{r,0}f_0)^{1.84} ,
\label{eq:fitke}
\end{equation}
and
\begin{equation}
\dot{M}_{\rm sat} = 7.86\times 10^{-12} M_{\odot}{\rm yr^{-1}}(B_{r,0}f_0)^{1.62}.
\label{eq:fitmdot}
\end{equation}
(See \S \ref{sec:extgen} for more general expressions with dependences on 
on $r_0$ and $M$ dependences.)
Both relations give the similar dependences on $(B_{r,0}f_0)$. 

The comparison between Equations (\ref{eq:scke}) and (\ref{eq:fitke}) gives 
\begin{equation}
c_{\rm E}\langle \delta v_0^2\rangle|_{\rm sat} \propto (B_{r,0}f_0)^{0.84}
\label{eq:satsc}
\end{equation}
for the saturated cases, where we neglect the weak dependence on the 
photospheric density, $\rho_0$.
In order to understand this relation, let us start from the dependence 
of $c_{\rm E}$ on $\delta v_0^2$ in a fixed $(B_{r,0}f_0)$. 
In the unsaturated state with small $\langle \delta v_0^2\rangle$, an 
increase of $\langle \delta v_0^2\rangle$ enhances $c_{\rm E}$, because 
the reflection of the \Alfven waves is suppressed (\S \ref{sec:rlrtr}). 
When one further increases $\langle \delta v_0^2\rangle$  to the saturated 
state, $c_{\rm E}$ starts to decrease on account of the
radiation loss as discussed in this section and finally 
$c_{\rm E}\langle \delta v_0^2\rangle$ itself decreases. 
At the saturation, $c_{\rm E}\approx 0.05$, which is nearly independent from 
$(B_{r,0}f_0)$ (Figure \ref{fig:Fph_SWKE}). Equation (\ref{eq:satsc}) 
indicates that $\langle \delta v_0^2\rangle$ that gives the saturation 
of the stellar wind is roughly proportional to $(B_{r,0}f_0)$ (we assume 
$0.84\approx 1$.). 
In cases with large $(B_{r,0}f_0)$, large amplitude \Alfven waves 
can propagate to higher locations owing to larger $v_{\rm A}$ and accordingly 
the smaller nonlinearity of the \Alfven waves as discussed so far. 

\begin{figure}
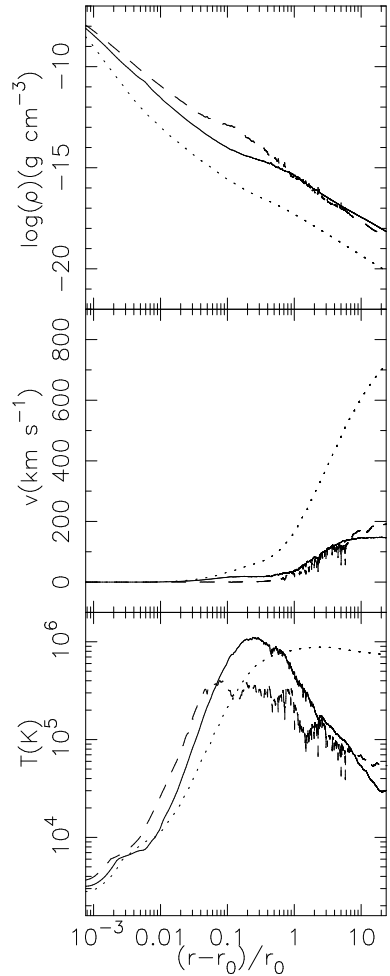

\begin{center}
\FigureFile(50mm,){Ysunstrave_cmp2_03.ps}
\end{center}
\caption{Comparison of the time-averaged wind structures of three cases 
tabulated in Table \ref{tab:sat1}. 
The less active case (Model Dd$+3\gamma$), the intermediate 
case (Model De$+4\gamma$), and the active case (Model Df$+5\gamma$) are shown 
in dotted, solid, and dashed lines. 
From top to bottom, the densities, radial velocities, and temperatures are
compared. } 
\label{fig:tmavstr2}
\end{figure}

Equations (\ref{eq:fitke}) and (\ref{eq:fitmdot}) are the maximum kinetic 
energy and mass loss rate 
for given $(B_{r,0}f_0)$. 
The standard case for the current solar wind (Model Cb$0\alpha$; 
Table \ref{tab:ref}) gives only 
8\% of the maximum kinetic energy ($L_{\rm K,out}/L_{\rm K,out,sat}=0.08$). 
Therefore, by increasing $\delta v_0$ at the photosphere, the stellar wind 
kinetic energy could be raised up to 12.5 times but cannot exceed it. 
In order to get further large kinetic energy, 
a flux tube with larger $(B_{r,0}f_0)$ is required. 

\begin{table*}
{\footnotesize 
\begin{tabular}{|c|c|c|c|c||c|c|c|c|c|c|}
\hline
Model & $\delta v_0$ & $B_{r,0}$(kG)$\times f_0$ & 
$h_{\rm l}$ ($r_0$) & $(L_{\rm A}f)_0$ & $(L_{\rm A}f)_{\rm tc}$ 
& $c_{\rm T}$ 
& $\dot{M}$ ($M_{\odot}$yr$^{-1}$) & $v_{r,{\rm out}}$ 
& $L_{\rm K,out}$ & $(L_{\rm R}f)_{\rm tc}$ \\
\hline
\hline
Dd$+3\gamma$ &  3.79 & ${4}/{1600}$ & 0.1 & $2.0\times 10^{30}$ 
& $1.2\times 10^{29}$ & 0.061 & $3.3\times 10^{-13}$ & $733$
& $5.6\times 10^{28}$ & $4.9\times 10^{27}$\\
\hline
De$+4\gamma$ & 5.35 & ${8}/{3200}$ & 0.1 & $3.9\times 10^{30}$ 
& $1.1\times 10^{30}$ & 0.28 & $6.0\times 10^{-12}$ & $145$
& $4.0\times 10^{28}$ & $4.7\times 10^{29}$ \\
\hline
Df$+5\gamma$ & 7.57 & ${16}/{6400}$ & 0.1 & $7.8\times 10^{30}$ 
& $1.6\times 10^{30}$ & 0.21 & $4.0\times 10^{-12}$ & $191$
& $4.6\times 10^{28}$ & $1.5\times 10^{30}$\\
\hline
\end{tabular}
\caption{Input parameters and outputs for the cases shown in Figure 
\ref{fig:tmavstr2}. The unit of $\delta v_0$ and $v_{r,{\rm out}}$ is 
km s$^{-1}$, and the unit of $L$ is erg s$^{-1}$. $c_{\rm T}$ is the surviving 
fraction of the \Alfven waves from the photosphere to the top of the 
chromosphere (Equation \ref{eq:trnsm}).
\label{tab:sat1}
}
}
\end{table*}

An important point regarding Equations (\ref{eq:fitke}) and (\ref{eq:fitmdot}) 
is that the saturation level is not determined by only $B_{r,0}$ but 
$B_{r,0}f_0$; open flux tubes need to occupy a sufficient fraction of the 
stellar surface ({\it i.e.}, large $f_0$) in addition to large field 
strength at the photosphere.  
This is also consistent with the explanation of the saturated 
stellar winds from the change of the magnetic topology that closed 
magnetic loop structure, which is expected to dominanty cover the 
surface of active stars, inhibits the stellar winds \citep{wod05}.  
A large fraction of closed loops corresponds to a small $f_0$, and then,  
stellar winds arising from such conditions will also become weak in our 
simulations. 

\subsubsection{Atmospheric Structure}

We further study the saturated state of the stellar winds by examining  
some simulated wind structures. Figure \ref{fig:tmavstr2} compares the  
three cases, summarized in Table \ref{tab:sat1}. 
We select runs with the same $B_{r,0}f_0 = 2.5$ G, 
but different $\delta v_0$. From the first case (Model Dd$+3\gamma$. 
dotted) 
through third case (Model Df$+5\gamma$, dashed), the input 
energy flux ($\propto B_{r,0}f_0\delta v_0^2$) from the photosphere 
increases by twice each.  

From the first case (Model Dd$+3\gamma$, dotted) to the 
second case (Model De$+4\gamma$, solid), the atmospheric 
structure drastically changes. 
The density of the second case is $\sim$10 times larger in the 
chromosphere and corona and $\sim$100 times larger in the stellar wind 
region. On the contraty, the wind speed becomes very slow because the dense 
wind cannot be effectively accelerated. 
As a result of the high density, 
the radiative loss ($(L_{\rm R}f)_{\rm tc}$ in Table. \ref{tab:sat1}) of the 
second case is nearly 100 times larger than that of the first case, even 
though the input wave energy is only twice. Although the mass loss rate 
($\dot{M}\propto \rho v_r$ in Table. \ref{tab:sat1}) is also larger due to 
the enhanced density, the kinetic energy luminosity 
($L_{\rm K,out}\propto \rho v_r^3$ in Table. \ref{tab:sat1}) slightly decreases 
because of the low velocity, which is recognized as the saturation 
of $L_{\rm K,out}$ in the top panel of Figure \ref{fig:Ftr_dep}.
While the maximum temperature of the second case is higher than that of the 
first case, the temperature decreases more rapidly because the thermal 
conduction is relatively ineffective against the adiabatic cooling 
in the denser corona.

The comparison between the second (Model De$+4\gamma$) and third 
(Model Df$+5\gamma$) cases exhibits an example of a saturated state. 
The overall density and velocity structures are quite similar. Then, 
both $\dot{M}$ and $L_{\rm K,out}$ from these two cases are similar; 
the mass loss rate also saturates at this stage. 
By a close look one can see that the third case gives slighly larger 
density from the chromosphere to the low coronal region, which leads to the 
$\sim 3$ times larger radiative loss as shown in Table. \ref{tab:sat1}. 
The maximum temperature of the third case is smaller than that of 
the second case because of the enhanced radiative loss.

\section{Discussions}
\label{sec:dis}
\subsection{Comparison with Wood et al.}
\label{sec:cmpobs}


\begin{figure}
\begin{center}
\FigureFile(84mm,){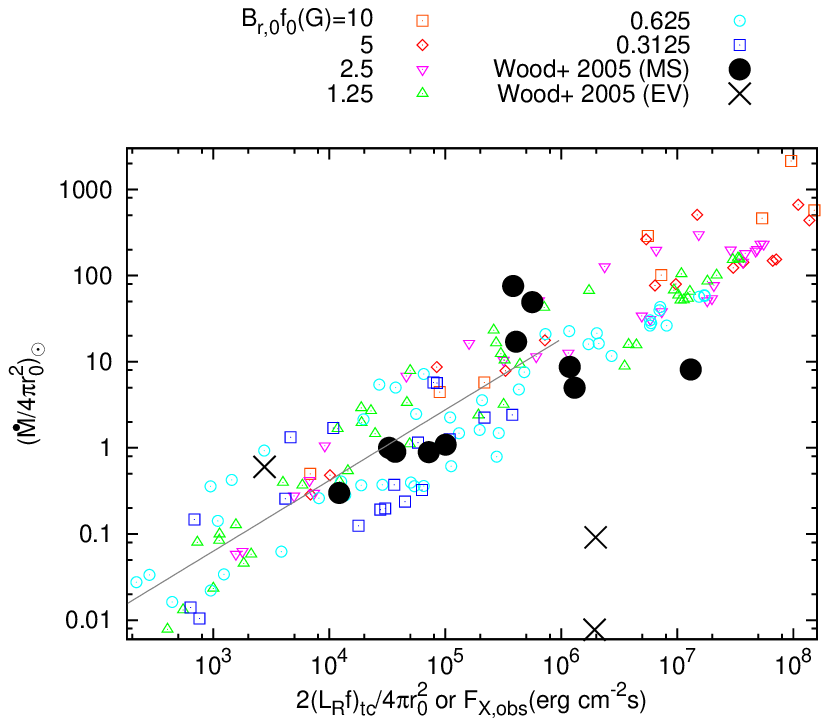}\\
\FigureFile(84mm,){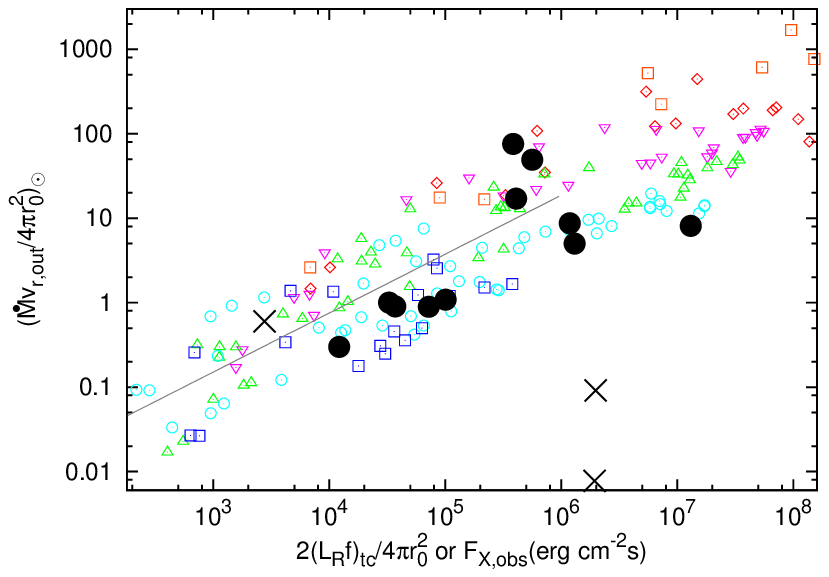}\\
\FigureFile(84mm,){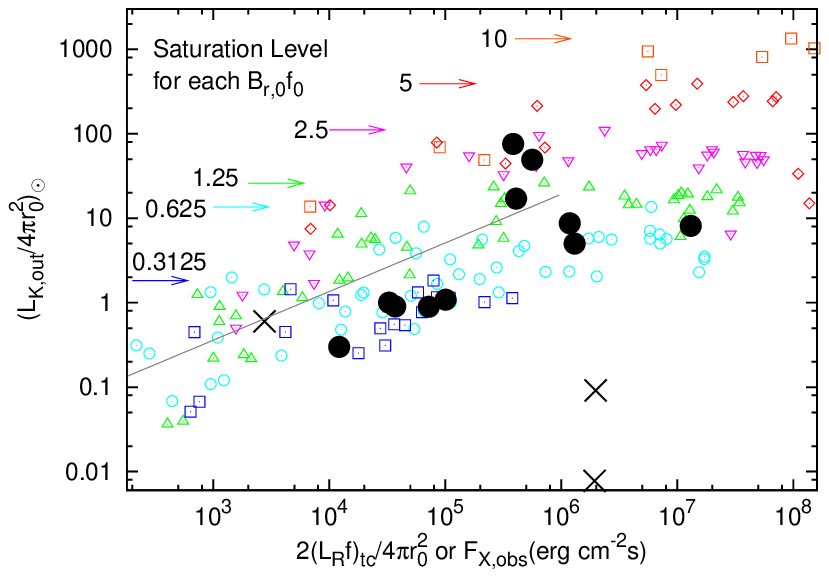}
\end{center}
\caption{Comparison of the simulation results (color symbols) with the 
observed data by Wood et al.(2002; 2005; big black symbols; circles 
indicate main sequence stars and crosses are evolved stars.)
The horizontal axis 
displays the X-ray flux (for Wood et al. data) or the radiation from open flux 
tube regions with $T\ge 2\times 10^4$ K multiplied by 
$c_{\rm r}=2$ (for our results; see {\it text} 
for detail). From top to bottom we compare the mass loss rates ($\propto 
\rho v_{r,{\rm out}}$), the ram pressures ($\propto 
\rho v_{r,{\rm out}}^2$), and the kinetic energy luminosities 
($\propto \rho v_{r,{\rm out}}^3$)  
divided by the stellar surface areas and normalized by the present-day solar 
level ($\dot{M} =2\times 10^{-14}M_{\odot}$yr$^{-1}$ and $v=400$ km s$^{-1}$).  
Then, the observational data points are plotted at the same 
relative positions 
in the three panels, because Wood et al (2002; 2005) adopt the constant 
$v=$400 km s$^{-1}$ for their hydrodynamical simulations to estimate 
the observational $\dot{M}$. 
The solid line in each panel is the power-law fit to the 
simulation data with $F_{\rm X}<10^6$ erg cm$^{-2}$s$^{-1}$ shown in 
Equation (\ref{eq:plfit}).
The color arrows in the bottom panel indicate the saturation levels of 
$L_{\rm K,out}$ for the respective $B_{r,0}f_0$. 
}
\label{fig:cmpobs}
\end{figure}

In this section, we compare our results with the results
by Wood et al. (2002; 2005). 
They derived the mass loss rates by comparing the observed 
Ly$\alpha$ absorptions of nearby G, M, K-type stars with the hydrodynamical 
simulations on the assumption of the spherical symmetry and the steady-state 
with a fixed terminal speed (=400 km s$^{-1}$) of 
the stellar winds. Although the derived mass loss rates might have 
uncertainties because of these assumptions, this work is the best effort 
to observationally determine the mass loss rates of the low-mass stars.

\subsubsection{Evolution with X-ray flux}
\label{sec:evxr}
As a key presentation in their papers, they plotted the mass loss
rates with the X-ray fluxes of the observed solar-type stars;  
we here put our results in the same diagram. 
The mass loss rates obtained in our simulations can be directly compared 
with the {results by Wood et al.(2002; 2005)}. On the other hand, 
the X-ray fluxes of our simulations considerably underestimate the actual
X-ray fluxes because our simulations only treat open flux tubes although 
it is expected that the observed X-rays are mainly from closed loops.
From an observational point of view, the UV flux from solar-type stars 
is well-correlated with the X-ray flux \citep{ayr97}. 
Based on these considerations, 
we use the radiation flux from the gas with $T>2\times 10^4$ K.

In order to compare our results with the observed X-ray fluxes, $F_{\rm X}$, we 
need to estimate a conversion factor, $c_{\rm r}$, from the radiation flux 
from the gas with $T\ge 2\times 10^4$ K in the simulated open flux tubes to 
the actual X-ray flux averaged over the entire stellar surface, 
where $c_{\rm r}$ is explicitly in an equation,   
\begin{equation}
F_{\rm X} = c_{\rm r} \frac{(L_{\rm R}f)_{\rm tc}}{4\pi r_0^2}. 
\end{equation}
Here, $\frac{(L_{\rm R}f)_{\rm tc}}{4\pi r_0^2}$ is the radiation flux 
from the gas with $T\ge 2\times 10^4$ K in the open flux tubes. 
While it does not include the X-ray from closed loops as discussed above, 
it includes the radiation in the UV range in addition to the X-ray; $c_{\rm r}$ 
could be either larger or smaller than unity. Because we do not have a 
quantitatively accurate way to estimate $c_{\rm r}$, we simply 
choose a value of $c_{\rm r}$ to give a reasonable fit to the observation.  
In Figure \ref{fig:cmpobs}, we show our results with $c_{\rm r}=2$ in comparison
with the observed data. 
From top to bottom, we compare the mass loss rates, $\dot{M}$, ram pressures, 
$\dot{M}v_{r,{\rm out}}$, and kinetic energy luminosities, $\dot{M}
\frac{v_{r,{\rm out}}^2}{2}$, 
divided by the surface areas of the stars and normalized by the current 
solar level. 
The three panels essentially compare $\rho v_{r,{\rm out}}$, 
$\rho v_{r,{\rm out}}^2$, and $\rho v_{r,{\rm out}}^3$.
We use $\dot{M}_\odot=2\times 10^{-14} M_{\odot}$ yr$^{-1}$ and 
$v_{\odot}=400$ km s$^{-1}$ as the normalizations in Figure 
\ref{fig:cmpobs} according to Wood et al. (2002; 2005). 
Since they assume the same wind speed $=400$ km s$^{-1}$ for the  
different stars, 
we put the observed data 
points at the same relative positions in the three panels. 

One can recognize that our simulations exhibit more clear saturations in 
$\rho v_{r,{\rm out}}^2$ and $\rho v_{r,{\rm out}}^3$. 
On the other hand, $\rho v_{r,{\rm out}}$ 
shows an increasing trend rather than a saturation. 
This is mainly because active cases with large radiation fluxes give 
systematically slower wind velocities (Fig.\ref{fig:tmavstr1}).
The Ly$\alpha$ absorptions in the hydrogen walls of 
astrospheres are correlated with the ram pressures of stellar winds 
\citep{wl98}. In this sense, the ram pressure (the middle panel) is a more 
physically plausible variable. 
In this panel the simulation results explains the observed distribution 
quite well except for the evolved stars. 

\citet{wod05} obtained $\dot{M}\propto F_{\rm X}^{1.34\pm 0.18}$ from their 
data of the ``unsaturated'' stars. On the other hand, a theoretical model 
by \citet{hj07} gives a shallower dependence, $\dot{M}\propto F_{\rm X}^{0.5}$.
We perform power-law fits to our simulation data with 
$F_{\rm X}\le 10^6$ erg cm$^{-2}$s$^{-1}$ corresponding to the unsaturated 
cases for the three panels of Figure \ref{fig:cmpobs}: 
{\color{black}
\begin{eqnarray}
\hspace{-0.6cm}
\frac{\dot{M}}{4\pi r_0^2} &=& \frac{\dot{M}_{\odot}}{4\pi R_{\odot}^2}
\left(\frac{F_{\rm X}}{2.9\times 10^4{\rm erg\;cm^{-2}s^{-1}}}\right)^{0.82} 
\nonumber \\
\hspace{-0.6cm}
\frac{\dot{M}}{4\pi r_0^2}v_{\rm out} &=& \frac{\dot{M}_{\odot}}{4\pi R_{\odot}^2}
v_{\odot}\left(\frac{F_{\rm X}}{1.5 \times 10^4{\rm erg\;cm^{-2}s^{-1}}
}\right)^{0.70} \nonumber \\
\hspace{-0.6cm}
&=&\frac{\dot{M}_{\odot}}{4\pi R_{\odot}^2} v_{630}
\left(\frac{F_{\rm X}}{2.9 \times 10^4{\rm erg\;cm^{-2}s^{-1}}
}\right)^{0.70} \label{eq:plfit}\\
\hspace{-0.6cm}
\frac{\dot{M}}{4\pi r_0^2}v_{\rm out}^2
&=& \frac{\dot{M}_{\odot}}{4\pi R_{\odot}^2}v^2_{\odot}
\left(\frac{F_{\rm X}}{5.9 \times 10^3{\rm erg\;cm^{-2}s^{-1}}
}\right)^{0.58} \nonumber \\
\hspace{-0.6cm}
&=& \frac{\dot{M}_{\odot}}{4\pi R_{\odot}^2}v_{630}^2
\left(\frac{F_{\rm X}}{2.9 \times 10^4
{\rm erg\;cm^{-2}s^{-1}}}\right)^{0.58}, \nonumber
\end{eqnarray}
}
where again the normalizations are 
$\dot{M}_\odot=2\times 10^{-14} M_{\odot}$ yr$^{-1}$ and 
$v_{\odot}=400$ km s$^{-1}$.
Most of the unsaturated cases with smaller inputs $(L_{\rm A}f)_0$ of our 
simulations give faster wind speeds than $v_{\odot}$. Then, we are also showing 
the relations of $\dot{M}v_{\rm out}$ and $\dot{M}v_{\rm out}^2$ with 
the normalization of {\color{black}$v_{630}=630$ km s$^{-1}$} 
for the wind speeds instead of 
$v_{\odot}$, which gives the exactly same normalization of 
$F_{\rm X}=2.9\times 10^4$ erg cm$^{-2}$s$^{-1}$ as that for $\dot{M}$. 
The value for the velocity normalization is interestingly close to 
the escape velocity, $v_{\rm esc,0}=618$ km s$^{-1}$.
The above scaling relations show that 
as multiplied by $v_{\rm out}$ once and twice, the power-law index 
decreases. This is because the stellar wind becomes slower as the wind 
density ($\sim \rho_{\rm tc}$) increases with increasing $(L_{\rm A}f)_0$ 
(\S \ref{sec:pwcon}). Interestingly enough, the obtained power-law index 
falls between the values by \citet{wod05} and by \citet{hj07}. 

The top panel as well as the middle and bottom panels of 
Figure \ref{fig:cmpobs} further show {\color{black}horizontal} scatters of 
the simulation data that exceed one order of magnitude. This indicates 
that {\color{black}the X-ray fluxes could vary even though the mass loss 
rates are unchanged,} which actually occur during the 11-year cycle of the sun.

\subsubsection{Time Evolution --Speculative Scenario--}
\label{sec:tmevl}

\begin{figure*}
\begin{center}
\FigureFile(170mm,){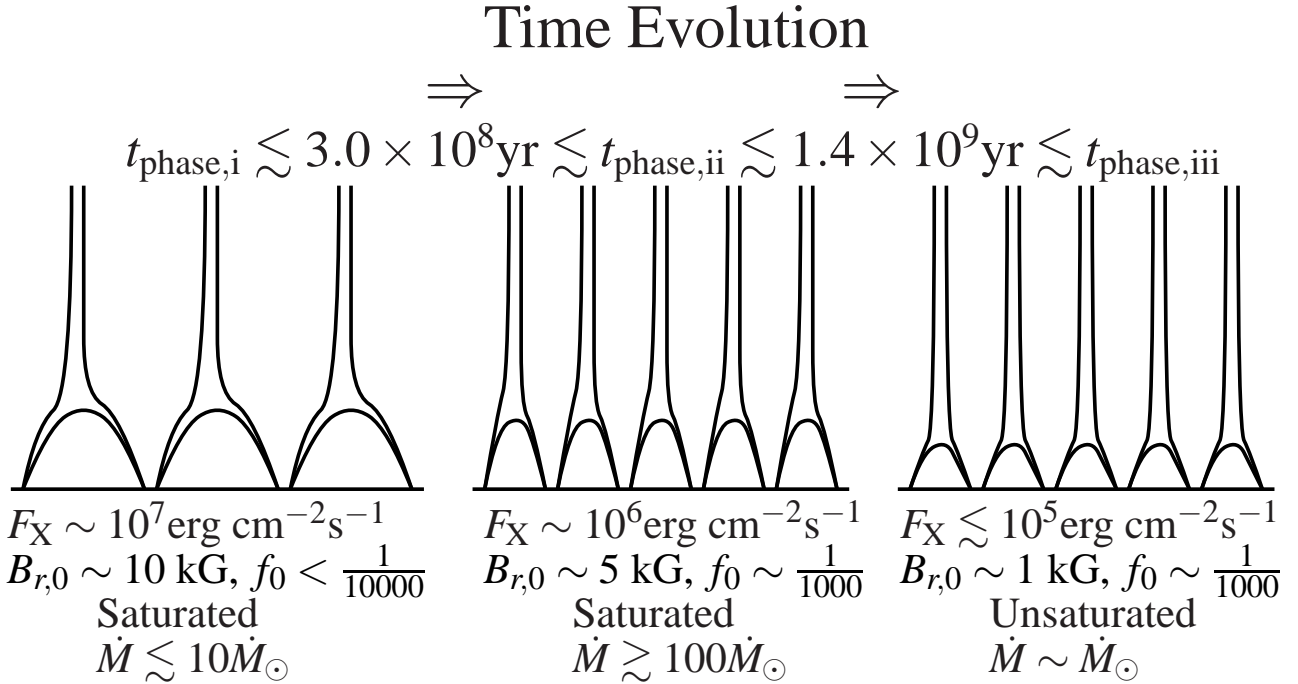}
\end{center}
\caption{Speculative scenario of the time evolution of properties 
of open flux tubes based on  Wood et al. (2002; 2005). 
(see \S \ref{sec:tmevl} for details.)}
\label{fig:sctev}
\end{figure*}

In realistic situations, the average $B_{r,0}f_0$ of the open flux tubes 
of a sun-like star would change with time.  
Comparison of the observed values with the simulation results in 
the middle or bottom panels of Figure \ref{fig:cmpobs} enables us to 
raise a following speculative scenario on the time evolution of the stellar 
wind (Figure \ref{fig:sctev}). 
{\color{black} In Figure \ref{fig:cmpobs} stars generally evolve from the 
right (larger $F_{\rm X}$) to the left (smaller $F_{\rm X}$). }
The most active star among the observed data at 
$F_{\rm X}=1.3\times 10^7$erg cm$^{-2}$s$^{-1}$ is well explained by a small 
value of $B_{r,0}f_0 < 1$ G. 
Probably, this star is mostly covered by closed 
loops and the filling factor of open flux tubes is very small, say $f_0 
\lesssim 1/10000$. Then, even if the photospheric field strength is strong, 
e.g. $B_{r,0}\sim 10$ kG, $B_{r,0}f_0$ becomes small $<1$ G, which gives 
the low saturation level of the stellar wind. 
With the stellar evolution, the magnetic topology is relaxed so that 
$f_0$ increases to $\sim 1/1000$. At this stage with 
$F_{\rm X}\sim 10^6$erg cm$^{-2}$s$^{-1}$, the stellar wind flux 
increases owing to the increase of $f_0$, although 
stars are still in the saturated state. In the subsequent evolution, 
$B_{r,0}$ decreases with time and stars are deviated from the saturated state. 
The stellar wind kinetic energies eventually settle down to the current 
solar level.  

We should be very careful when comparing the evolution of magnetic fields 
in this scenario with observed magnetic fields in young active stars 
(e.g. Saar 2001; Donati \& Landstreet 2009), 
because observationally determined $B_{r,0}$ and $f_0$ sensitively depend on 
the spatial resolution. Even on the sun, kilo-Gauss patches 
in coronal holes are a recent new discovery by the high resolution 
observation by SOT/HINODE \citep{tsu08}. Observations with lower resolutions
would give smaller $B_{r,0}$ as a result of spatial smoothing, and accordingly 
the derived $f_0$ would be larger, because $B_{r,0}f_0$ is independent from 
the spatial resolution. As discussed previously, the contribution from 
close magnetic loops, of which the strength also depends on the spatial 
resolution, needs to be taken into account.  

In this scenario, an increase of $f_0$, namely the change of the magnetic 
topology, plays an important role in the evolution of the mass loss from 
sun-like stars. In this context, it is similar to the consideration raised 
by \citet{wod05} in which closed magnetic structures that cover the surface 
of very active stars inhibit the stellar winds but gradual opening of 
these closed fields liberate the stellar winds streaming out. 
In our scenario, however, the radiation loss which is enhanced by the 
increase of the density in the atmospheres is as important in the saturation 
of the stellar winds as the change of the magnetic topology.

We can infer the time evolution in a little more quantitative way by using 
$F_{\rm X}$ as a tracer of stellar ages. X-ray observations of sun-like 
stars give a negative correlation between X-ray luminosity, $L_{\rm X}$, 
and stellar age, $t_{\rm age}$, \citep{mag87}, with some of them are 
indirectly connected through stellar rotation periods \citep{gud97}. 
\citet{gud07} derived a relation, 
\begin{equation}
L_{\rm X} \approx (3\pm 1) \times 10^{28}\left(\frac{t}{10^9{\rm yr}}
\right)^{-1.5\pm 0.3} \; {\rm erg\;s^{-1}}, 
\label{eq:Lxtage}
\end{equation}
for sun-like stars. 
Following the speculated scenario in Figure \ref{fig:sctev}, we classify 
the evolution into the three phases in terms of $L_{\rm X}=4\pi 
r_0^2 F_{\rm X}$ estimated from our simulations: 
\begin{enumerate}[{(}i{):}]
\item{saturated \& weak wind phase: \\ 
$F_{\rm X}\gtrsim 3.0\times 10^6$ erg cm$^{-2}$s$^{-1}$ or $L_{\rm X}\gtrsim 
1.8\times 10^{29}$ erg s$^{-1}$\\
$t_{\rm phase,i}\lesssim 3.0\times 10^8$ yr}
\item{saturated \& strong wind phase: \\
$3.0\times 10^5$ erg cm$^{-2}$s$^{-1}$ $\lesssim F_{\rm X}\lesssim 
3.0\times 10^6$ erg cm$^{-2}$s$^{-1}$ or $1.8\times 10^{28}$ erg s$^{-1}$ 
$\lesssim L_{\rm X}\lesssim 2\times 10^{29}$ erg s$^{-1}$ \\
$3.0\times 10^8$ yr $\lesssim t_{\rm phase,ii} 
\lesssim 1.4\times 10^9$ yr}
\item{unsaturated wind phase: \\
$F_{\rm X}\lesssim 3.0\times 10^5$ erg cm$^{-2}$s$^{-1}$ or $L_{\rm X}\lesssim 
1.8\times 10^{28}$ erg s$^{-1}$\\
$t_{\rm phase,iii}\gtrsim 1.4\times 10^9$ yr
}
\end{enumerate}
Here, the timescale of each phase is estimated by using Equation 
(\ref{eq:Lxtage}).

Combining Equations (\ref{eq:plfit}) and (\ref{eq:Lxtage}), we can further 
derive the scalings for the stellar winds as a function of time for the 
unsaturated cases of our simulations: 
{\color{black}
\begin{eqnarray}
\frac{\dot{M}}{4\pi r_0^2} &=& 1.6\times \frac{\dot{M}_{\odot}}{4\pi R_{\odot}^2}
\left(\frac{t}{t_{\odot}}\right)^{-1.23}, \nonumber\\
\frac{\dot{M}}{4\pi r_0^2} v_{\rm out} &=& 1.5\times \frac{\dot{M}_{\odot}}
{4\pi R_{\odot}^2}v_{630}
\left(\frac{t}{t_{\odot}}\right)^{-1.05}, \label{eq:mdot-t}\\
\frac{\dot{M}}{4\pi r_0^2}v_{\rm out}^2 &=& 1.4\times \frac{\dot{M}_{\odot}}
{4\pi R_{\odot}^2}v_{630}^2 
\left(\frac{t}{t_{\odot}}\right)^{-0.87}, \nonumber
\end{eqnarray}
}
where $t_{\odot}=4.6\times 10^9$ yr. 
{\color{black}These time dependences are less steep than the relation 
obtained in \citet{wod05}, $\dot{M}\propto t^{-2.33\pm 0.55}$. 
This is mainly because our derived relation of $\dot{M}$ on $F_{\rm X}$ 
(Equation \ref{eq:plfit}) is shallower than the relation obtained 
in \citet{wod05}, $\dot{M}\propto F_{\rm X}^{1.34\pm 0.18}$. Also, \citet{wod05} 
adopted a different relation of $F_{\rm X}\propto t^{-1.74\pm 0.34}$ 
\citep{ayr97} from the relation we are using (Equation \ref{eq:Lxtage}),
which further contributes to the shallower dependence of 
Equation (\ref{eq:mdot-t}). }


\subsection{Faint Young Sun Paradox}
\label{sec:fysp}
The luminosity of sun-like stars gradually increases during the
main sequence phase. Based on the standard stellar evolution calculation, 
the luminosity of the sun is 20-30\% smaller at early times \citep{gou81}. 
As a result, the surface temperature on the earth is expected to be below
the freezing temperature before $\sim$ 2 billion years ago. 
On the other hand, liquid water already existed at that time on the
earth as well as on the mars \citep{sm72,ke86,feu11}, 
which looks contradictory, known as the `faint young Sun paradox'. 
A high level of green house effects \citep{owe79,kas97}, 
and modification of the solar model \citep{whi95,mm07} are the two major 
possibilities which can solve the problem. 
We here discuss effects of the strong mass loss of the Sun \citep{wil87,sac03}. 

As already discussed in \S \ref{sec:rgrtr}, 
the saturation level of the stellar winds is determined by $B_{r,0}f_0$. 
The top panel of Figure \ref{fig:Ftr_dep} and Figure \ref{fig:cmpobs} 
indicate that stars with large $B_{r,0} f_0$ ($\sim$ 5-10G) could drive 
$\sim 500-1000$ times stronger mass loss with the X-ray flux in 
the observed range ($F_{\rm X}\le 2\times 10^7$ erg cm$^{-2}$s$^{-1}$), 
although such a large mass loss rate has not been observed so far. 
If such a strong mass loss actually continues $\sim$ 1 billion years 
as discussed in \S \ref{sec:tmevl} and Figure \ref{fig:sctev}, 
it contributes to the decrease of stellar masses. 

1000 times of $\dot{M}$ during 1 billion years gives the mass loss of 
0.02$M_{\odot}$.  If this is true in the Sun, the initial mass is 
2 \% larger than the present Sun. An increase of the stellar mass 
influences the radiation flux on planets around a star in two ways. 
First, the stellar radiation 
itself is larger; in sun-like stars, the luminosity, $L$, can be scaled 
by $L\propto M^{4.75}$ \citep{kw90,whi95}. A 2\% larger $M$ gives 
$\approx 10$ \% larger $L$. 
Second, the orbital radius, $a_{\rm p}$, of a planet shrinks 
in a manner to keep $M a_{\rm p}$ constant \citep{whi95,mm07}. 
Then, the radiation flux ($\propto 1/a_{\rm p}^2$) on the planet increases 
by $\approx 4$ \%. As the sum of these two effects, the radiation flux 
increases $\approx 15$ \%, 
which could considerably compensate the early faint sun and 
avoid the freezing temperature on the earth.  

In this paper, we are focusing on the quasi-steady component of the 
stellar winds from open flux tubes during the main sequence phase. 
In active sun-like stars, 
coronal mass ejections (CMEs), which probably involves magnetic reconnections 
of closed loops (e.g. Ohyama \& Shibata), might significantly contribute to 
the mass losses. \citet{aar12} estimated mass loss rates by CMEs from 
pre-main sequence stars by extrapolating the observed relation between 
X-ray flare energy and CME mass in the present Sun. Although their main
target is not CMEs during early main sequence phases, 
which is crucial in the faint young Sun paradox, we can infer by a simple 
interpolation that the relative contribution of CMEs to the total mass loss 
is larger at earlier times than the present contribution ($\sim 10$\%). 

\subsection{Extended Variable Chromosphere}
\label{sec:evchr}

\begin{figure}
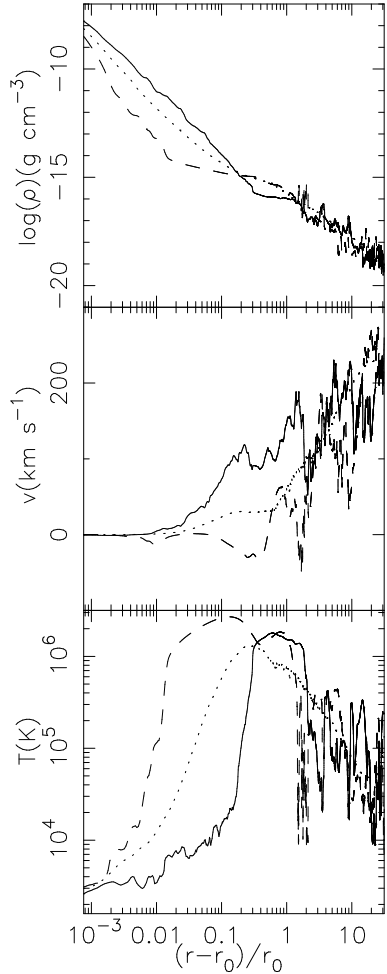

\begin{center}
\FigureFile(50mm,){Ysunstrave-snp_cmp1.ps}
\end{center}
\caption{Comparison of snap-shot wind structures at 
$t=258.2$ hr (solid) and 
$t=269.6$ hr (dashed) for Model Dd$+4\gamma$. 
The dotted lines are the time-average between 
$t=92.45$ hr and $369.7$ hr.}
\label{fig:snpstr1}
\end{figure}

We have discussed in \S \ref{sec:rlrtr} that, when increasing the input 
wave energy from the current solar level, the chromospheric materials are 
lifted by the magnetic pressure of the \Alfven waves (\S \ref{sec:refstr} and 
Figure \ref{fig:chdens}). 
An extended chromosphere is a universal 
feature in our simulations with relatively large energy inputs.  
Another important aspect 
is that the atmospheres behave more dynamically. 
In particular, the transition regions move 
up and down because of the thermally unstable region ($T\gtrsim 10^5$ K) 
of the radiative cooling function \citep{lm90,sd93}.  
Figure \ref{fig:snpstr1} shows snapshot structures (solid and dashed lines) 
of an active case (Model Dd$+4\gamma$) with $\delta v_0=5.35$ km s$^{-1}$, 
$B_{r,0}=4$ kG, $f_0=1/1600$, and $h_{\rm l}=0.1 r_0$, in comparison with the 
time-averaged wind structure (dotted lines). In the time-averaged structure 
the temperature gradually increases from $T=10^4$ K at 
$r\approx 0.01 r_0$ to $T=10^6$ K at $r\approx 0.15 r_0$. However, one 
can recognize from the snapshots that this is simply 
a result of the long-time average. 
The snapshot structures at the two different times show that the sharp 
transition region moves up and down; at 
$t=258.2$ hr (solid lines) the chromosphere is extended 
up to $r\approx 0.2 r_0$, while at 
$t=269.6$ hr (dashed lines) 
the corona is getting down to $r\approx 0.01 r_0$ with a part of dense 
coronal gas falling down \citep{pin09}.   
 
\begin{figure}
\begin{center}
\FigureFile(90mm,){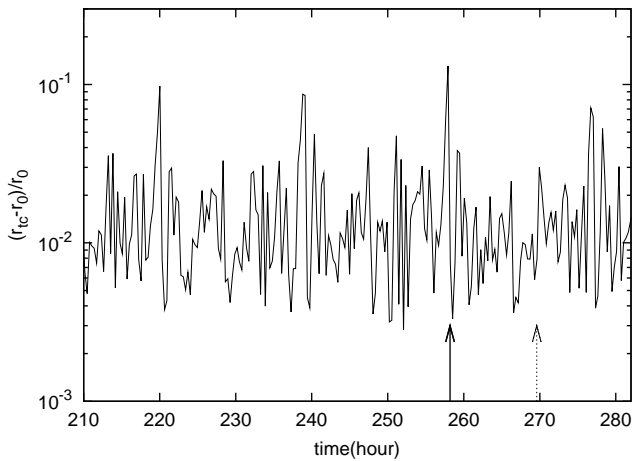}
\end{center}
\caption{Time evolution of the top of the chromosphere, 
$r_{\rm tc}$, of Model Dd$+4\gamma$. 
The times at which the snap-shot structures 
are displayed in Figure \ref{fig:snpstr1} are indicated by the arrows.} 
\label{fig:dchrtev}
\end{figure}

\begin{figure}
\begin{center}
\FigureFile(90mm,){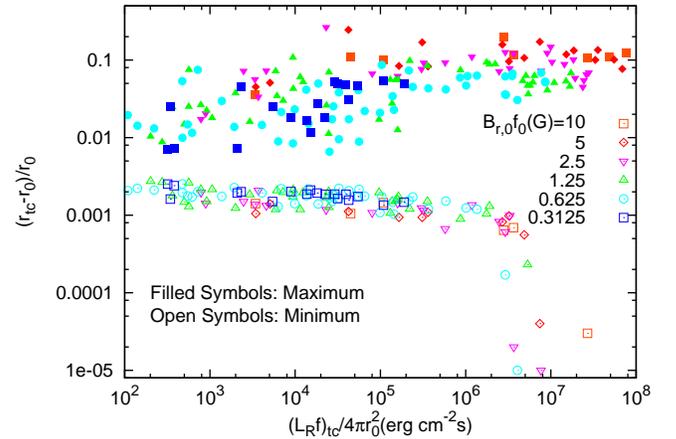}
\end{center}
\caption{Location of the top of the chromosphere, $r_{\rm tc}$, on 
radiation flux of each run. 
The filled symbols distributed in the upper half of the panel indicate 
the maximum heights during the simulations, and the open symbols near the 
bottom of the panel indicate the minimum heights. In some cases with large 
radiation flux, $(L_{\rm R}f)_{\rm tc}/4\pi r_0^2$, the minimum of $r_{\rm tc}$ 
is below the displayed region in the panel. }
\label{fig:extchr}
\end{figure}

Figure \ref{fig:dchrtev} displays the time evolution of the top of 
the chromosphere ($=$the base of the transition region), $r_{\rm tc}$, 
at $T=2\times 10^4$ K, of Model Dd$+4\gamma$.
The figure clearly shows that the transition region moves up and down 
dynamically. Within the displayed duration of 72 hours ($=$3 days), 
the chromosphere extends to $\gtrsim 1.1 r_0$ by 3 times. On the other 
hand, its motion is sometimes quite fast with the timescale less than an hour; 
for instance, from $t=250$ hr to 253 hr, the transition region moves 
from $1.003 r_0$ to $1.05 r_0$ with the timescale of 10 minutes. 
We checked the Fourier transformation of the time-sequence data in 
Figure \ref{fig:dchrtev} to find a roughly flat spectrum without any distinct 
features (not shown). This indicates that the motion of the transition 
region does not show clear characteristic timescales but is stochastic 
with the timescale from minutes to a day. 
This is because the timescale is controlled by the multiple processes of 
the wave heating, the radiative cooling, and the thermal conduction. 
The timescale of the wave heating is an order of minutes, which corresponds 
to the periods of the injecting perturbations from the photosphere. 
On the other hand, the timescale of the radiative cooling and the thermal 
conduction could be large depending on the background physical condition 
\citep{sy02}. 
For instance, the timescale, $\tau_{\rm c}$, of the thermal conduction can 
be estimated 
\citep{sy02} as 
\begin{equation}
\tau_{\rm c} \approx 3 \left(\frac{\rho_{\rm tr}}{10^{-15}{\rm g\; cm^{-3}}}
\right)\left(\frac{\Delta r_{\rm tr}}{0.1 r_0}\right)^2\left(
\frac{T}{10^6{\rm K}}\right)^{-5/2} {\rm hr},
\end{equation}
where $\rho_{\rm tr}$ is the density at the transition region and 
$\Delta r_{\rm tr}$ is the width of the transition region.
The equation shows that $\tau_{\rm c}$ sensitively depends on 
$\Delta r_{\rm tr}$.
For the typical condition of the present-day sun, $\Delta r_{\rm tr}<0.01r_0$, 
which gives $\tau_{\rm c}<20$ min. In active stars, {\rm e.g.} Model 
Dd$+4\gamma$ in Figure \ref{fig:snpstr1}, $\Delta r_{\rm tr}$ could be 
large enough to give $\tau_{\rm c}\sim$ hours -- a day.

Recently, \citet{cze12} observed the atmosphere of a planet-hosting young 
star, CoRoT-2a, by using the technic of the Rossiter-McLaughlin effect. 
They reported that it could have a very thick chromosphere 
with height of 10-20\% of the radius, which is much larger than 
the chromosphere of the present Sun which is only up to $<$ 1\% 
of the radius. Moreover, they observed an asymmetry in the chromospheric 
Rossiter-McLaughlin effect, which might arise from inhomogeneities of 
the chromospheric structure.

Figure \ref{fig:extchr} shows the height of the chromosphere of 
each simulation run.  In the figure, we plot both the maximum (filled symbols) 
and minimum (open symbols) values of $r_{\rm tc}$ at $T=2\times 10^4$ K during 
the time evolution of each simulation.  
In cases with small radiation flux, the differences between 
the maximum and minimum heights are small and $r_{\rm tc}$ is typically 
$0.01 r_0$, which is consistent with the present-day Sun 
(e.g. Imada et al.2011).
As the radiation flux increases, the maximum height increases and the 
minimum height decreases, namely in active cases the transition regions 
largely move up and down.  
The minimum heights in cases with relatively large 
radiation flux (in the right part of the figure) are considerably small, 
$r_{\rm tc}-r_0 < 10^{-4}r_0$($=70$ km). Moreover in some cases the minimum 
heights are in the outside of the displayed range. In these cases, the 
temperature increases to $T>2\times 10^4$ K from the next mesh 
($\approx 10^{-5}r_0 =$ 7 km) to the photosphere at some instants.  

\subsection{General Scalings of $L_{\rm K,out}$ \& $\dot{M}$}
\label{sec:extgen}

In our simulations, we consider the star with the same stellar parameters as 
the sun, $M=M_{\odot}$, $r_0=R_{\odot}$, $T_{\rm eff}=5780$ K, and 
$\rho_0=10^{-7}$ g cm$^{-3}$. 
Then, the scaling relations for $L_{\rm K,out}$ and $\dot{M}$ in Equations 
(\ref{eq:scke}) and (\ref{eq:mssc}) have been derived for the fixed 
stellar parameters. 
Leaving the basic stellar parameters as variables, we can derive more general 
expressions for $L_{\rm K,out}$ and $\dot{M}$:
\begin{eqnarray}
& &L_{\rm K,out} = c_{\rm E} (L_{\rm A}f)_0 = c_{\rm E}\Phi_{\rm B} 
\sqrt{\frac{\rho_0}{4\pi}} \langle \delta v_0^2\rangle  \nonumber \\
&=& 2.1\times 10^{27}{\rm erg\;s^{-1}}\left(\frac{c_{\rm E}}{0.017}\right)
\left(\frac{r_0}{R_{\odot}}\right)^2 
\left(\frac{\rho_0}{10^{-7}{\rm g\; cm^{-3}}}\right)^{1/2}\nonumber \\
& &
\left(\frac{B_{r,0}f_0}{1.25{\rm G}}\right)
\left\langle\left(\frac{\delta v_0}
{1.34{\rm km\; s^{-1}}}\right)^2\right\rangle,
\label{eq:sckegen}
\end{eqnarray}
and 
\begin{eqnarray}
\dot{M} &=& c_{\rm M}\frac{(L_{\rm A}f)_0 r_0}{GM}
=c_{\rm M}\Phi_{\rm B} \sqrt{\frac{\rho_0}{4\pi}}
\frac{\langle \delta v_0^2\rangle r_0}{GM}, \nonumber \\
&=& 2.2\times 10^{-14}M_{\odot}{\rm yr}^{-1}\left(\frac{c_{\rm M}}{0.023}\right)
\left(\frac{r_0}{R_{\odot}}\right)^3
\left(\frac{M}{M_{\odot}}\right)^{-1}\nonumber \\
& & \hspace{-0.5cm}
\left(\frac{\rho_0}{10^{-7}{\rm g\; cm^{-3}}}\right)^{1/2}
\left(\frac{B_{r,0}f_0}{1.25{\rm G}}\right)
\left\langle\left(\frac{\delta v_0}
{1.34{\rm km\; s^{-1}}}\right)^2\right\rangle, 
\label{eq:msscgen}
\end{eqnarray}
where the relation for $\dot{M}$ is derived in a similar manner 
to the Reimers (1975) formula 
(see also Schr\"{o}der \& Cuntz 2005). 

These relations should be tested by numerical simulations with stars 
with different masses, radii, and photospheric densities, which is our 
future work. It should be noted that 
the variables in Equations (\ref{eq:sckegen}) and (\ref{eq:msscgen}) 
are not independent each other. 
For example, limiting to low-mass ($M\lesssim M_{\odot}$) main 
sequence stars, $r_0$ is related to $M$, $r_0\propto M^{0.8}$  
(\S 22 of Kippenhahn \& Weigert 1990).
Also, the density, $\rho_0$, at a photosphere can be estimated from the 
stellar basic parameters. For instance, in the cool star condition with 
effective temperature, 4000 K$\lesssim T_{\rm eff}\lesssim$ 6000 K, 
$\rho_0 \propto \left(\frac{M}{r_0^2}\right)^{0.6}T_{\rm eff}^{-3}$ 
(\S 9 of Gray 1992; Suzuki 2007). Using these dependences and $T_{\rm eff} 
\propto M^{0.4}$ (\S 22 of Kippenhahn \& Weigert 1990), we can replace the 
dependences on $r_0$ and $\rho_0$ in Equations (\ref{eq:sckegen})  and 
(\ref{eq:msscgen}) with the dependences on $M$:
\begin{eqnarray}
L_{\rm K,out} &=& 2.1\times 10^{27}{\rm erg\;s^{-1}}
\left(\frac{c_{\rm E}}{0.017}\right)
\left(\frac{M}{M_{\odot}}\right)^{0.82} \nonumber \\
& &
\left(\frac{B_{r,0}f_0}{1.25{\rm G}}\right)
\left\langle\left(\frac{\delta v_0}
{1.34{\rm km\; s^{-1}}}\right)^2\right\rangle,
\label{eq:sckegen2}
\end{eqnarray}
and 
\begin{eqnarray}
\dot{M} 
&=& 2.2\times 10^{-14}M_{\odot}{\rm yr}^{-1}\left(\frac{c_{\rm M}}{0.023}\right)
\left(\frac{M}{M_{\odot}}\right)^{0.62}\nonumber \\
& & 
\left(\frac{B_{r,0}f_0}{1.25{\rm G}}\right)
\left\langle\left(\frac{\delta v_0}
{1.34{\rm km\; s^{-1}}}\right)^2\right\rangle, 
\label{eq:msscgen2}
\end{eqnarray}

Equation (\ref{eq:msscgen}) or (\ref{eq:msscgen2}) could be a general 
formula to estimate mass loss rates of stars with a surface convection zone 
(see also Cranmer \& Saar 2011).
Once the basic stellar 
parameters, $r_0$, $M$, and $T_{\rm eff}$, are obtained, Equation 
(\ref{eq:msscgen}) or (\ref{eq:msscgen2}) has the three undetermined 
parameters, $c_{\rm M}$, 
$(B_{r,0}f_0)$, and $\delta v_0$. One can use a typical value, $c_{\rm M}\approx 
0.02$ from the simulation results. 
$\delta v_0$ is expected to be a fraction of the sound speed at the 
surface; $\delta v_0=1.34$ km s$^{-1}$ of the standard case corresponds 
to $\sim 20$ \% of the sound speed. $(B_{r,0}f_0)$, magnetic field strength in 
{\it open field regions}, is the most unknown parameter among the three.   
If $\dot{M}$ is obtained by other methods, Equation (\ref{eq:msscgen}) 
or (\ref{eq:msscgen2}) 
can be used to derive $B_{r,0}f_0$ in the opposite way.

We also show general formulae for the saturated values, 
Equations (\ref{eq:fitke}) and (\ref{eq:fitmdot}), with taking into 
account the dependences on $M$ and $r_0$: 
\begin{eqnarray}
L_{\rm K,out,sat} &=& 2.05\times 10^{28}{\rm erg\;s^{-1}}(B_{r,0}f_0)^{1.84} 
\left(\frac{r_0}{R_{\odot}}\right)^2, \nonumber \\
&=& 2.05\times 10^{28}{\rm erg\;s^{-1}}(B_{r,0}f_0)^{1.84} 
\left(\frac{M}{M_{\odot}}\right)^{1.6},
\label{eq:fitkegen}
\end{eqnarray}
and
\begin{eqnarray}
\hspace{-0.6cm}\dot{M}_{\rm sat} 
&=& 7.86\times 10^{-12} M_{\odot}{\rm yr^{-1}}(B_{r,0}f_0)^{1.62}
\left(\frac{r_0}{R_{\odot}}\right)^3 \left(\frac{M}{M_{\odot}}\right)^{-1}, 
\nonumber \\
&=& 7.86\times 10^{-12} M_{\odot}{\rm yr^{-1}}(B_{r,0}f_0)^{1.62}
\left(\frac{M}{M_{\odot}}\right)^{1.4},
\label{eq:fitmdotgen}
\end{eqnarray}
where we here again use $r_0\propto M^{0.8}$ for low-mass main sequence 
stars.

\subsection{Stellar Rotation}
\label{sec:strot}
It is expected that stellar rotation is an important ingredient in 
determining stellar magnetic activities \citep{sku72,ayr97,gud97,lin12}. 
Probably related the steep differential rotation owing to the fast rotation 
of young active stars \citep{hy11}, the magnetic field is effectively 
amplified in fast rotating solar-type stars \citep{god12}.
We take into account the effects of the strong magnetic fields on the stellar 
winds by incorporating the wide ranges of the parameters on the magnetic field. 

In addition to the generated magnetic field strength, the stellar rotation 
directly affects the dynamics of stellar winds through the magnetocentrifugal 
force \citep{wd67,coh10}. 
The angular momentum of a star is outwardly transported by the magnetorotating 
wind, which is the main cause of the deceleration of the stellar rotation 
\citep{mat12}. We do not take into account this effect in our simulations.
We here briefly discuss possible roles of the magnetocentrifugal force 
in the dynamics of the stellar winds. 

In general, stronger magnetic field and faster rotation affects the dynamics 
of stellar winds. Magnetic energy per mass in magnetorotating 
winds (Weber \& Davis 1967; see also \S 9 of Lamers \& Cassinelli 1999) is 
written as 
\begin{equation}
\epsilon_{\rm B} = -\frac{r\Omega B_r B_\phi}{4\pi\rho v_r} 
\approx \frac{(r\Omega)^2 B_r^2}{4\pi\rho v_r^2},  
\label{eq:mgrtpf}
\end{equation} 
where $\Omega$ is the rotation frequency of a star, $B_\phi$ is the 
azimuthal component of the background magnetic field ($B_r B_\phi<0$ because 
of winding field lines), 
and we here consider the problem in the outer region where the super-radial 
expansion of the flux tubes already finishes, $f=1$. For the second
transformation in Equation (\ref{eq:mgrtpf}), we have used the relation 
for the geometry of magnetic field lines under the steady-state condition, 
\begin{equation}
\frac{B_{\phi}}{B_r} = \frac{v_{\phi}-r\Omega}{v_r} \approx 
-\frac{r\Omega}{v_r}, 
\end{equation} 
where $v_{\phi}$ is the azimuthal component of velocity. 
If $\epsilon_{\rm B}$ is larger than kinetic energy per mass, 
the magnetocentrifugal force dominantly works and the wind velocity 
will be faster than obtained in our simulations. 
Then, we can use a nondimensional parameter, 
\begin{equation}
C_{B\Omega}=\frac{\epsilon_{\rm B}}{\rho v_r^2/2}=\frac{r^4B_r^2\Omega^2} 
{4\pi r^2 \rho v_r^4/2} = \frac{(B_{r,0}f_0 r_0^2)^2\Omega^2}{\dot{M}v_r^3/2}, 
\label{eq:cbo1}
\end{equation}
which measures the importance of magnetocentrifugal force\footnote{
What we are doing here is essentially the comparison between
the Michel velocity, $v_{\rm M}$, \citep{mic69} and the escape velocity, 
$v_{\rm esc,0}$. Neglecting numerical factors with an order of unity, 
$C_{B\Omega}=1$ corresponds to $v_{\rm M}\approx v_{\rm esc,0}$.}.
The effect of magnetocentrifugal force becomes important in proportion 
to magnetic energy and rotation energy but in inversely proportion to 
a mass loss rate. 

We can estimate $C_{B\Omega}$ from a typical case of our simulations for 
active stars, 
\begin{eqnarray}
\hspace{-0.3cm}C_{B\Omega}=0.57\left(\frac{B_{r,0}f_0}{5{\rm G}}\right)^2
\left(\frac{M}{M_{\odot}}\right)\left(\frac{r_0}{R_{\odot}}\right)
\left(\frac{\dot{M}}{100\dot{M}_{\odot}}\right)^{-1} \nonumber \\
\left(\frac{v_{r,{\rm out}}}{400{\rm km\;s^{-1}}}\right)^{-3}
\left(\frac{\Omega}{0.1\Omega_{\rm K}}\right)^2, 
\label{eq:cbo2}
\end{eqnarray}
where $\dot{M}_{\odot} = 2\times 10^{-14}M_{\odot}$yr$^{-1}$ is the mass loss 
rate of the current Sun, and $\Omega_{\rm K}=\sqrt{\frac{GM}{r_0^3}}$ is 
the Kepler (breakup) velocity at the stellar surface. Equation (\ref{eq:cbo2}) 
is normalized by 10 \% of $\Omega_{\rm K}$, which gives the dependences on 
$M$ and $r_0$. The rotation of the current Sun is $\Omega_{\odot}\approx 
0.005 \Omega_{\rm K}$. Then, Equation (\ref{eq:cbo1}) indicates that the effect 
of the magnetocentrifugal force is small if the rotation frequency of a star 
which gives 100 times of $\dot{M}$ is smaller than 20 times of the current 
solar rotation frequency. In stars with rotation period $\lesssim$ 1 day 
the magnetocentrifugal acceleration is important and the terminal 
velocity becomes faster, whereas it also depends on the mass loss rate.

\section{Summary}
We have investigated how the properties of the stellar winds from 
sun-like stars are determined from the magnetic fields and velocity 
perturbations at the photosphere. 
We performed the 163 models of the MHD simulations in the range of 
$\approx$ 4 orders of magnitude of the \Alfven wave energy flux from the 
photosphere. We examined the properties of the stellar winds of 
these simulations mainly from an energetics point of view.  
0.1-10\%, with the typical value, $\sim 1$\%, of the input Poynting flux 
of the \Alfven waves from the photosphere is finally transported to the 
kinetic energy of the stellar winds. We derived the scaling relations 
that dermine the kinetic energy luminosity and mass loss rate of the stellar 
winds from the properties at the stellar photosphere (Equations \ref{eq:scke}, 
\ref{eq:mssc}, \ref{eq:sckegen} -- \ref{eq:msscgen2}). 
These relations can be used to estimate the mass losses 
from solar-type stars \citep{hj07,cs11}. 

We can classify the simulated stellar winds into the unsaturated regime 
and the saturated regime. 
When the input wave energies are small, the stellar winds are in the 
unsaturated state; an increase of the input wave energy directly leads 
to strong stellar winds. 
In this regime, the reflection of \Alfven waves 
dominantly controls the energetics and dynamics of the stellar winds. 
In the chromosphere and low corona, the density rapidly 
decreases with height. Consequently, the \Alfven speed rapidly increases, 
which makes \Alfven waves effectively reflected because of the deformation 
of the wave shapes. The reflection is more severe for a smaller wave energy 
input because the density decreases more rapidly without the support from 
the magnetic pressure ($\propto \delta B_{\perp}^2/8\pi$) associated with 
the \Alfven waves. In extreme cases, 
more than 99\% of the input Poynting energies are reflected back before 
reaching the top of the chromosphere. The mass loss rates of these cases 
are as small as $\sim$ 1\% of that of the current sun.

With increasing the input wave energy, the density decreases more slowly in 
the chromosphere owing to the support by the magnetic pressure of the 
\Alfven waves. The chromosphere of the active cases is time-dependently 
extended to 10 -- 20 \% of the radius, and the transition region moves up 
and down (\S \ref{sec:evchr}) mainly because of the thermally unstable region 
($T\gtrsim 10^5$ K) of the radiative cooling function of the gas with the 
solar abundance \citep{sd93}.  
According to the slow decrease of the density in the chromospheres, 
the \Alfven speed changes more gradually, which suppresses the reflection 
of the \Alfven waves. 
When one increases the input wave energy, the kinetic energy and 
the mass loss rate of the stellar winds increase more rapidly than 
the linear dependence on the input energy. 
In cases with the large energy inputs for active stars, more than 10\% of the 
input wave energy from the photosphere reach the top of the chromosphere. 

For the large wave energy inputs, the stellar winds are in the saturated 
state; an increase of the input energy does not raise the kinetic 
energy of the winds but enhances the radiative loss. Increasing the input 
wave energy from the unsaturated state to the saturated state, the densities 
of the transition region and corona monotonically increase, which gives 
larger kinetic energy and mass loss of the stellar winds. 
The increase of the densities also raise the radiative cooling. Since the 
radiative loss is proportional to $\rho^2$ in the optically thin plasma, 
the increase of the radiative loss is faster than the increase of the 
wind kinetic energy with increasing the input wave energy from the photosphere. 
In the saturated state, most of the Poynting energy at the top of 
the chromospheres is transferred to the radition in the transition regions 
and coronae, and a tiny fraction of the energy is available for the 
kinetic energy of the stellar winds.

The saturation level has the power-law dependence on $B_{r,0}f_0$, 
$L_{\rm K,out,sat}\propto (B_{r,0}f_0)^{1.82}$ (Equation \ref{eq:fitke}), 
which can be understood from the nonlinear dissipation of the \Alfven waves. 
$B_{r,0}f_0$ determines the magnetic field strength in the upper corona 
and wind region where the super-radial expansion of the flux tubes already 
finishes. For small $B_{r,0}f_0$, the nonlinearity of the \Alfven waves, 
$\delta B_{\perp}/B_r$ becomes systematically larger, which results in 
faster dissipation. Then, more wave energy dissipates at lower altitudes 
where the density is higher, and mostly escapes by the radiation loss.  
For large $B_{r,0}f_0$, the opposite explanation can be done; more wave energy 
remains up to higher altitudes and contributes to driving the strong stellar 
winds.   

The positive correlation of the saturation level with $B_{r,0}f_0$ 
indicates that in order to drive strong stellar winds the filling factor 
of open flux tubes should not be so small, as well as the surface magnetic 
field should be sufficiently strong. This is consistent with the explanation 
that closed magnetic loops inhibit stellar winds from active stars. 
Based on our results, we introduce a speculative scenario for 
the evolution of the stellar winds from sun-like stars (\S \ref{sec:cmpobs}; 
Figure \ref{fig:sctev}). 
At very early time, the surface of a star is mostly covered with closed 
magnetic field (small $f_0$), and then, the mass loss rate is not so large 
because of the low saturation level, even though the surface field 
strength ($B_{r,0}$) is large. 
The magnetic field structure is eventually relaxed so that open 
flux tubes occupy a larger fraction ($f_0$ increases) and the mass loss rate 
increases. After that, the mass loss rate  gradually decreases to 
the current solar level with decreasing magnetic field strength.

\bigskip
This work was supported in part by Grants-in-Aid for 
 Scientific Research from the MEXT of Japan, 22864006.
We are grateful to the anonymous referee for many constructive 
suggestions to improve the paper. We also thank Drs. Brian Wood and 
Jeffery Linsky for many fruitful discussions.
Numerical simulations in this work were partly carried out at the Yukawa 
Institute Computer Facility, SR16000.

\end{document}